\documentclass[numberedappendix,12pt,preprint]{emulateapj}
\slugcomment{submitted to ApJ}  

\usepackage[]{amsmath}
\usepackage{longtable}
\usepackage{lscape}
\usepackage{multirow}
\usepackage{ccaption}
\usepackage{array}

\shorttitle{Abell 2744 from HFF and GLASS}
\shortauthors{Wang et al. (2015)}

\usepackage{color}			            
\definecolor{midgray}{gray}{0.4}		
\definecolor{orange}{rgb}{1,0.5,0}      

\usepackage[colorlinks=true,citecolor=midgray,linkcolor=midgray]{hyperref}        

\newcommand{\simgt}{\,\rlap{\lower 3.5 pt \hbox{$\mathchar \sim$}} \raise 1pt \hbox {$>$}\,}
\newcommand{\simlt}{\,\rlap{\lower 3.5 pt \hbox{$\mathchar \sim$}} \raise 1pt \hbox {$<$}\,}

\newcommand{\NimgTOT}{179}   
\newcommand{\NimgUSE}{72}   
\newcommand{\NimgELtot}{7}  
\newcommand{\NsysELtot}{5}   
\newcommand{\NimgELhiQ}{5}   
\newcommand{\NELobjsQhi}{55}    
\newcommand{\NELQtwo}{18}       
\newcommand{\NELQthree}{16}     
\newcommand{\NELQfour}{34}      

\newcommand{\HST}{\emph{HST}}
\newcommand{\Spitzer}{\emph{Spitzer}}

\newcommand{\AB}{Abell~2744}
\newcommand{\cgs}{ergs s$^{-1}$cm$^{-2}$}

\begin{document}


\title{The Grism Lens-Amplified Survey from Space (GLASS). IV. Mass reconstruction of the lensing cluster Abell 2744 from frontier 
field imaging and GLASS spectroscopy}


\author{
X.~Wang$^{1}$, A.~Hoag$^2$, K.-H.~Huang$^2$, T.~Treu$^{3}$,  M.~Brada\v{c}$^2$, K.~B.~Schmidt$^1$, G.~B.~Brammer$^4$,
B.~Vulcani$^{5}$, T.~A.~Jones$^{1}$, R.~E.~Ryan,~Jr.$^{4}$, R.~Amor\'{i}n$^6$, M.~Castellano$^6$, A.~Fontana$^6$, E.~Merlin$^6$, 
M.~Trenti$^7$}
\affil{$^1$ Department of Physics, University of California, Santa Barbara, CA, 93106-9530, USA}
\affil{$^2$ Department of Physics, University of California, Davis, CA, 95616, USA}
\affil{$^3$ Department of Physics and Astronomy, UCLA, Los Angeles, CA, USA 90095-1547}
\affil{$^4$ Space Telescope Science Institute, 3700 San Martin Drive, Baltimore, MD, 21218, USA}
\affil{$^5$ Kavli Institute for the Physics and Mathematics of the Universe (WPI), The University of Tokyo Institutes for Advanced 
Study (UTIAS), the University of Tokyo, Kashiwa, 277-8582, Japan}
\affil{$^6$ INAF - Osservatorio Astronomico di Roma Via Frascati 33 - 00040 Monte Porzio Catone, I}
\affil{$^7$ School of Physics, The University of Melbourne, VIC 3010, Australia}
\email{xinwang@physics.ucsb.edu}

\begin{abstract}
We present a strong and weak lensing reconstruction of the massive cluster \AB, the first cluster for which deep \emph{Hubble Frontier 
Field} (HFF) images and spectroscopy from the \emph{Grism Lens-Amplified Survey from Space} (GLASS) are available. By performing a 
targeted search for emission lines in multiply imaged sources using the GLASS spectra, we obtain \NimgELhiQ{} high-confidence spectroscopic 
redshifts and 2 tentative ones. We confirm 1 strongly lensed system by detecting the same emission lines in all 3 multiple images.  
We also search for additional line emitters blindly and use the full GLASS spectroscopic catalog to test reliability of 
photometric redshifts for faint line emitters. We see a reasonable agreement between our photometric and spectroscopic redshift 
measurements, when including nebular emission in photometric redshift estimations. We introduce a stringent procedure to identify 
only secure multiple image sets based on colors, morphology, and spectroscopy. By combining 7 multiple image systems with secure 
spectroscopic redshifts (at 5 distinct redshift planes) with 18 multiple image systems with secure photometric redshifts, we 
reconstruct the gravitational potential of the cluster pixellated on an adaptive grid, using a total of \NimgUSE{} images.
The resulting mass map is compared with a stellar mass map obtained from the deep \emph{Spitzer} Frontier Fields data to study the 
relative distribution of stars and dark matter in the cluster. We find that the stellar to total mass ratio varies substantially 
across the cluster field, suggesting that stars do not trace exactly the total mass in this interacting system. The maps of 
convergence, shear, and magnification are made available in the standard HFF format.
\end{abstract}



\section{Introduction}
\label{sec:intro}

In the past two decades, gravitational lensing by clusters of galaxies has transitioned from an exotic curiosity to an invaluable 
tool for astrophysics and cosmology \cite[e.g.][]{K+N11}. Clusters can act as natural telescopes, magnifying background sources so 
that they appear brighter and more extended to the observer \cite[e.g.][]{Yee++96,Pet++02,Bradac:2012p28826,Shi++14,Bay++14}.  The 
gravitational lensing effect can also be used to reconstruct the spatial distribution of mass in the clusters themselves, thus 
shedding light on the physics of dark matter and structure formation 
\citep[e.g.][]{clowe06,Bra++06,San++08,New++13,Sha++14,2015ApJ...806....4M}.

In the past two years, clusters of galaxies as tools for cosmology and
astrophysics have become a major focus of a Hubble Space Telescope
(\HST) initiative. As part of the Hubble Frontier Field (HFF) program,
six clusters of galaxies and six parallel fields are being imaged to
unprecedented depths in seven optical and near-infrared (NIR) bands, using
the Wide Field Camera 3 (WFC3) and the Advanced Camera for Surveys
(ACS) \citep{2015ApJ...800...84C}. Similarly to previous public campaigns in
legacy fields such as the Hubble Deep Fields
\citep{Wil++96,Ferguson:2000p22537}, this major \HST{} campaign has
triggered coordinated observations with all major facilities, ranging
from Director's Discretionary time on the \Spitzer{} Space Telescope to
deep observations with the Chandra X-ray Telescope and many ground
based facilities (see the HFF website at
\url{http://www.stsci.edu/hst/campaigns/frontier-fields/FF-Data} for
details). 


One of the major drivers of the HFF initiative is to search for
magnified objects behind galaxy clusters. Accurate magnification maps
are needed to determine the unlensed (intrinsic) properties of these
galaxies. Fortunately, the deep images increase significantly the
number of known multiply imaged systems per cluster, thus improving
dramatically the fidelity of the mass models, and magnification
maps. Typical CLASH\footnote{\url{http://www.stsci.edu/~postman/CLASH/Home.html}}-like depth imaging of clusters
\citep[limiting magnitude $\sim$ 27 ABmag for a 5-$\sigma$ point
source][]{Postman:2012p27556} yields a handful of multiple image
systems per cluster \cite[e.g.][]{Joh++14}, however at the depth of the HFF
we expected to find tens of multiply image systems. This was
beautifully confirmed by the first papers that analyzed the HFF
imaging datasets
\citep[e.g.][]{2014arXiv1409.8663J,Ish++15,Lap++15}.

The spectacular increase in the quality of imaging data has
not yet been matched by advances in spectroscopic redshift ($z_{\textrm{spec}}$)
determination for the multiple images. Thus, modelers often have to
rely on photometric redshifts to incorporate multiple images in their
analysis, or sometimes they decide to leave the source redshift as a
free parameter to be inferred by the lens model itself. Whereas one
of these two choices is inevitable in the absence of other data, it is
also fraught with peril. Photometric redshifts can be very uncertain
and prone to catastrophic errors for sources that are well beyond the
spectroscopic limit, such as most of the faint arcs and
arclets. Similarly, letting the mass model determine the redshift of
the multiple images can potentially introduce confirmation bias in the
modeling process.

Obtaining spectroscopic redshifts for as many multiple image systems
as possible is thus a fundamental step if we want to improve the mass
models of the HFF clusters, and thus make the most of this
groundbreaking initiative. Several efforts are currently underway to
secure these redshifts, including our own based on the Grism Lens
Amplified Survey from Space (GLASS) data, which is presented in this paper. GLASS
is a large \HST{} program that has just completed obtaining deep spectroscopy in the fields of ten clusters,
including all six in the HFF program.  A full description of the
survey and its scientific drivers is given in paper I (Treu et
al. 2015, submitted).

In this paper we present a new mass model for the galaxy cluster
\AB, the first cluster for which both GLASS and HFF complete
datasets are available. By exploiting the exquisite imaging and
spectroscopic data we carry out a rigorous selection of multiple
images used to constrain the mass model. We use spectroscopic redshifts
-- including \NELobjsQhi{} new line emitters detected at high-confidence -- when available to supplement the
photometric redshift ($z_{\textrm{phot}}$) calibration. We then compare the inferred two
dimensional mass distribution to the distribution of stellar mass as
determined from the analysis of deep \Spitzer{} imaging data, indicating that the stellar to mass ratio
varies substantially across the cluster, which is expected given the merging nature of this cluster. 

The paper is organized as follows. In Section~\ref{sec:data} we give
an overview of the data. In Section~\ref{sec:datareduction} we
describe the reduction and analysis of the GLASS data. In
Section~\ref{sec:mul} we detail our algorithm for selection of
multiple image systems. In Section~\ref{sec:mass} we present our mass
model and study the relative distribution of stellar and total mass. In
Section~\ref{sec:conc} we summarize our conclusions. Throughout this
paper we adopt a standard concordance cosmology with $\Omega_m=0.3$,
$\Omega_{\Lambda}=0.7$ and $h=0.7$. All magnitudes are given in the AB
system \citep{Oke74}.

\section{Data}
\label{sec:data}

Being one of the most extensively studied galaxy clusters, \AB{} is observed on a spectrum extending
from X-ray to radio bands \citep[e.g.][]{2004MNRAS.349..385K}. In this paper, we focus on the optical and NIR
imaging and spectroscopy data newly acquired with the Hubble and
\Spitzer{} Space Telescopes, as part of the GLASS program (\ref{sec:glass}) and
HFF initiative (\ref{sec:HFF} and \ref{sec:spitz}).

\subsection{The Grism Lens-Amplified Survey from Space}
\label{sec:glass}

The Grism Lens-Amplified Survey from Space\footnote{\url{http://glass.physics.ucsb.edu}} (GLASS, GO-13459, P.I. Treu) is observing 
the cores of 10 massive galaxy clusters with the \HST{} WFC3 NIR grisms G102 and G141 providing an uninterrupted 
wavelength coverage from 0.8$\mu$m to 1.7$\mu$m.  The slitless spectroscopy is distributed over 140 orbits of \HST{} time in cycle 
21. The last GLASS observations were taken in January 2015.  Amongst the 10 GLASS clusters, 6 are targeted by the HFF (see Section~\ref{sec:HFF}) 
and 8 by the Cluster Lensing And Supernova survey with Hubble \citep[CLASH; 
P.I. Postman,][]{Postman:2012p27556}. Prior to each grism exposure, imaging through either F105W or F140W is obtained to assist 
the extraction of the spectra and the modeling of contamination from nearby objects on the sky.  The total exposure time per 
cluster is 10 orbits in G102 (with either F105W or F140W) and 4 in G141 with F140W. Each cluster is observed at two position 
angles (P.A.s) approximately 90 degrees apart to facilitate deblending and extraction of the spectra.

In concert with the NIR observations on the cluster cores
the two parallel fields corresponding to the two P.A.s are observed
with ACS's F814W filter and G800L grism.  Each parallel field has a
total exposure time of 7 orbits.  These observations map the cluster
infall regions.

A key focus point of GLASS is the advancement and improvement of the
lens models of the 10 surveyed clusters.  This paper focuses on the
modeling of the first cluster in the GLASS survey with complete GLASS
spectroscopy and HFF imaging as described in
Section~\ref{sec:datareduction}.

\subsection{Hubble Frontier Fields}
\label{sec:HFF}

The Hubble Frontier Fields initiative\footnote{\url{http://www.stsci.edu/hst/campaigns/frontier-fields}} (HFF, P.I. Lotz) is a 
Director's Discretionary Time legacy program with \emph{HST} devoting 840 orbits of \emph{HST} time to acquire optical ACS and 
NIR WFC3 imaging of six of the strongest lensing galaxy clusters on the sky; Abell 370, Abell 2744, MACSJ 2129, MACSJ 0416, MACSJ 
0717, and MACSJ 1149. For a 5-$\sigma$ point source, the limiting magnitudes are roughly 29 ABmag in both the ACS/optical (F435W, 
F606W, F814W) and WFC3/IR filters (F105W, F125W, F140W, F160W).  All six HFF clusters are included in the GLASS sample described 
in Section~\ref{sec:glass}.  The program was initiated in \emph{HST} cycle 21 and is planned for completion in 2016. The first 
cluster to have complete HFF and GLASS data available is the cluster studied in this paper, namely \AB.

An important aspect of the HFF efforts has been the community efforts
to model the lensing clusters.  Prior to the start of the HFF
observations five independent groups (CATS team, Sharon et al.,
Merten, Zitrin et al., Williams et al., and Bradac et al.) provided
lens models of the HFF clusters, which have been made available
online\footnote{\url{http://archive.stsci.edu/prepds/frontier/lensmodels/}}.
In the remainder of this work we will use our own models from Bradac
et al., but we will make comparison with several of the other publicly
available HFF lens models of \AB.


\subsection{\textit{Spitzer} Frontier Fields}
\label{sec:spitz}

The \Spitzer{} Frontier Fields program\footnote{\url{http://ssc.spitzer.caltech.edu/warmmission/scheduling/approvedprograms/ddt/frontier/}} 
(P.I. Soifer) is a Director's Discretionary Time program that images all six
strong lensing galaxy clusters targeted by the HFF in both warm IRAC channels ($3.6$ and $4.5~\mu$m).  With the addition
of archival imaging from  {\it IRAC Lensing Survey}
(P.I. Egami) we reduced the data using the methodology employed by the {\it Spitzer UltRa Faint SUrvey Program}
(SURFS UP, P.I. Brada\v{c}, \citealp{2014ApJ...785..108B}). IRAC imaging reaches $\sim\!50$~hr depth on the
{\it primary} (\AB{} cluster) and {\it parallel} ($\sim\!6'$ to the
west) fields.  There are four {\it flanking} fields with $\sim\!25$~hr
depth (two to the north and two to the south) of the primary and
parallel fields. {The nominal depth for a 5-$\sigma$ point source can reach 26.6 ABmag at 3.6 $\mu$m and
26.0 ABmag at 4.5 $\mu$m, respectively. However this sensitivity might be compromised in cluster center due to blending with 
cluster members and the diffuse intra-cluster light (ICL).}

For this work, we focus on the primary field of \AB{} that we have
processed in a fashion very similar to that discussed in
\citet{2014ApJ...785..108B,2014ApJ...786L...4R,2015arXiv150402099H}.  In brief, we applied additional warm-mission column pulldown and 
automuxstripe corrections as provided
by the \Spitzer{} Science Center to the corrected basic-calibrated data
(cBCD) to improve image quality, particularly near bright stars.  We
process these files through a standard overlap correction to equalize
the sky backgrounds of the individual cBCDs.  We drizzle these
sky-corrected cBCDs to $0\farcs6$~pix$^{-1}$ with a
\texttt{DRIZ\_FAC}$=\!0.01$ using the standard MOPEX
software\footnote{\url{http://irsa.ipac.caltech.edu/data/SPITZER/docs/dataanalysistools/tools/mopex/}}.
As a final note, there are $\sim\!1800$~frames (with
\texttt{FRAMETIME}$\sim\!100$~s) per output pixel in the deep regions.

\section{GLASS Observation and Data Reduction}
\label{sec:datareduction}

The two P.A.s of GLASS data analyzed in this study were 
taken on August 22 and 23 2014 (P.A. = 135) 
and October 24 and 25 2014 (P.A. = 233), respectively. 
Prior to reducing the complete GLASS data each exposure was checked for elevated background from He Earth-glow 
\citep{Brammer:2014p34990} and removed, if necessary. The \AB{} data was favorably scheduled so only the August 23 reads were 
affected and thus removed by our reduction pipeline.

The resulting total exposure times for the individual grism observations are: G102\_PA135 10929 seconds,
G141\_PA135 4212 seconds,
G102\_PA233 10929 seconds, and
G141\_PA233 4312 seconds.
The corresponding exposure times for the direct GLASS imaging are:
F105W\_PA135 1068 seconds,
F140W\_PA135 1423 seconds,
F105W\_PA233 1068 seconds, and
F140W\_PA233 1423 seconds.

\begin{figure*}
\centering
\includegraphics[width=\textwidth]{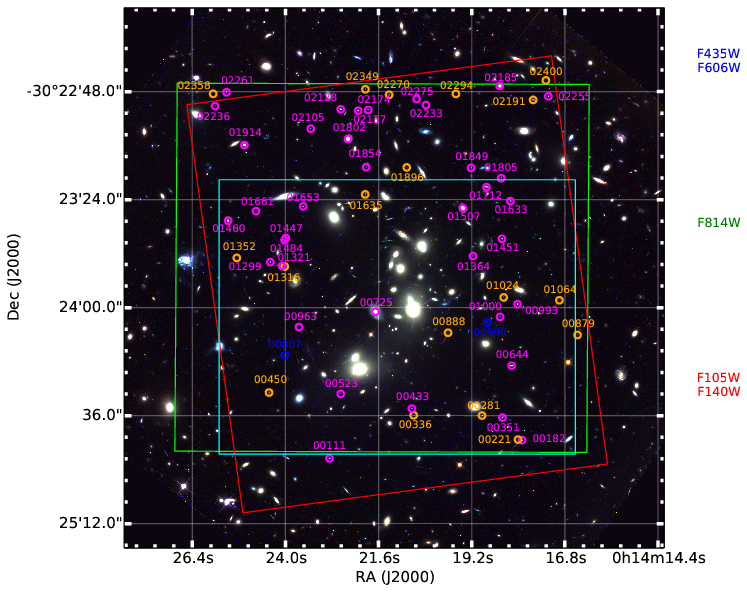}
\caption{The color composite image of \AB{} based on the HFF and GLASS imaging. The blue, green and red channels are composed by 
the filters on the right. The two distinct P.A.s of the spectroscopic GLASS pointings are shown by the red (P.A.=233 degrees) and 
green (P.A.=135 degrees) squares. Figure~\ref{fig:grisms} shows the grism images corresponding to these two P.A.s. The locations 
of the emission line objects from Table~\ref{tab:ELtot} are marked by circles, with color coding reflecting the GLASS 
spectroscopic redshift quality (cf. Section~\ref{subsec:targeted} and column ``Quality'' in Table~\ref{tab:ELtot}; 2=blue, 
3=orange, and 4=magenta). The cyan square shows the outline of Figure~\ref{fig:arcs_image}.}
\label{fig:image}
\end{figure*}

In Figure~\ref{fig:image} we show a color composite image of \AB,
using the optical and NIR imaging from the HFF combined with the NIR imaging from GLASS.
The red and green squares show the positions of the spectroscopic GLASS \AB{} field-of-views.

\begin{figure*}
\centering
\includegraphics[width=0.49\textwidth]{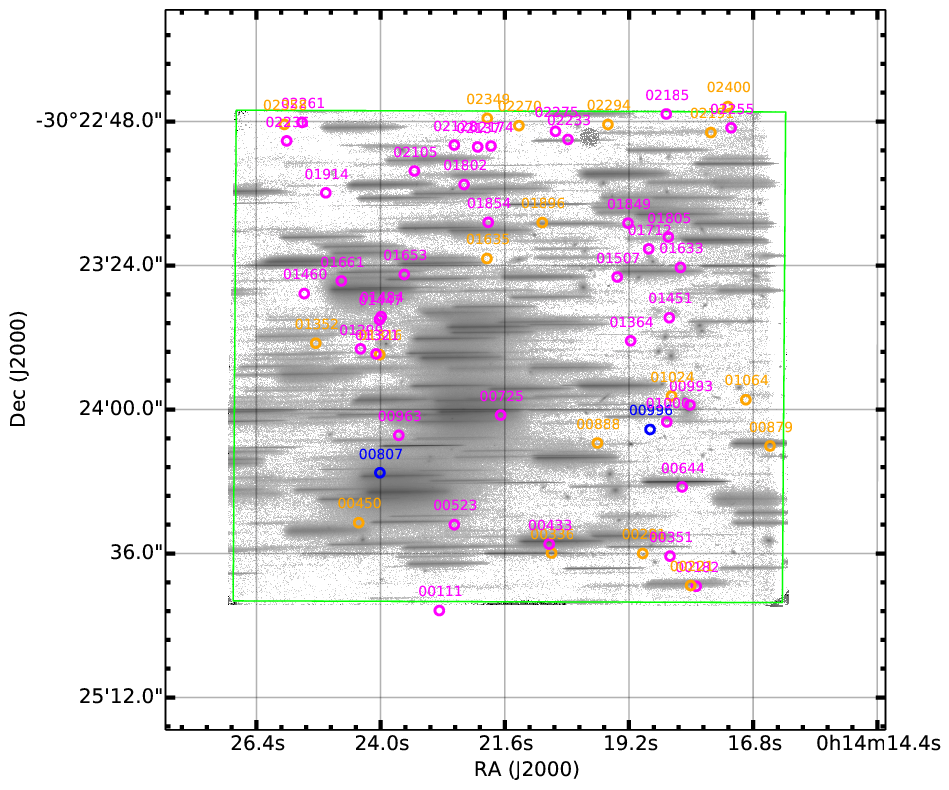}
\includegraphics[width=0.49\textwidth]{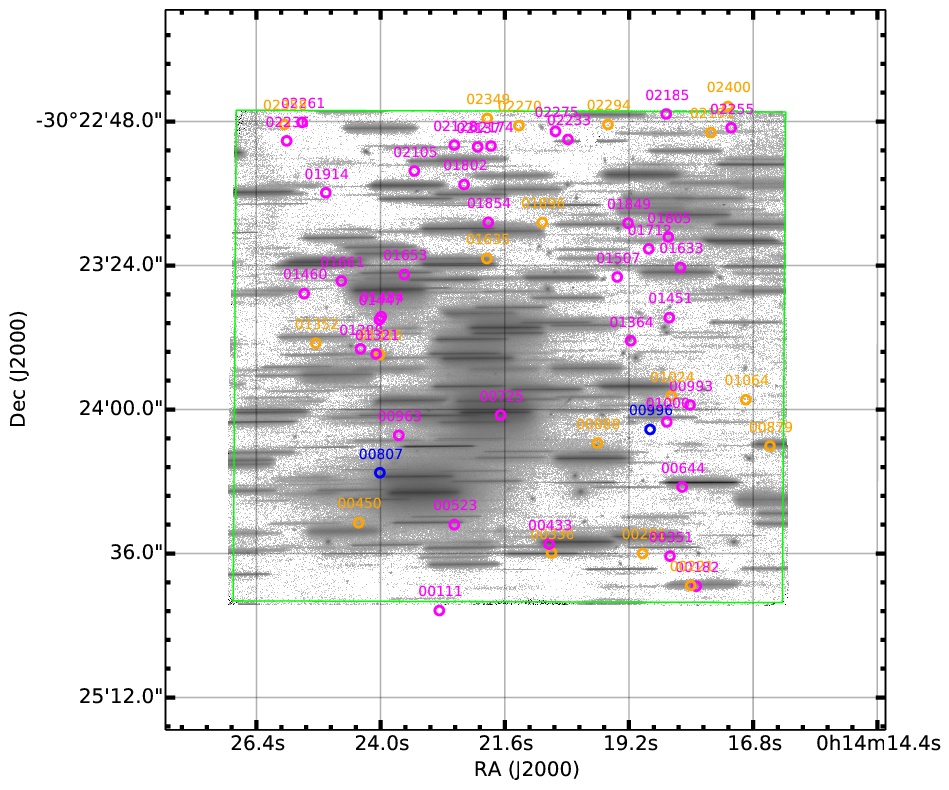}\\
\includegraphics[width=0.49\textwidth]{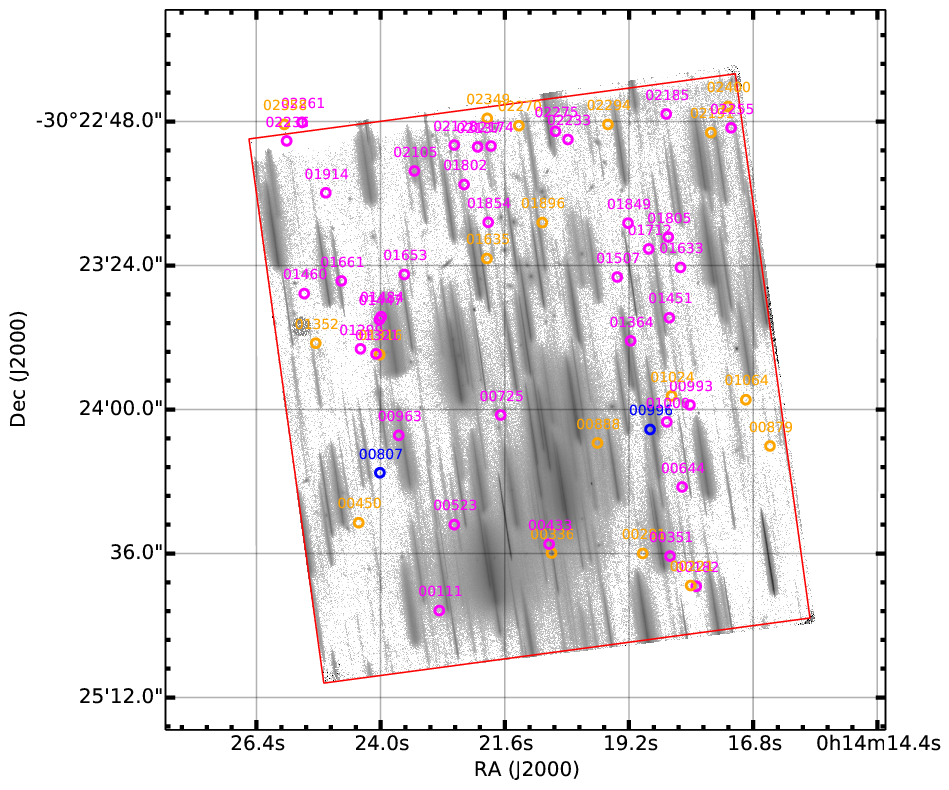}
\includegraphics[width=0.49\textwidth]{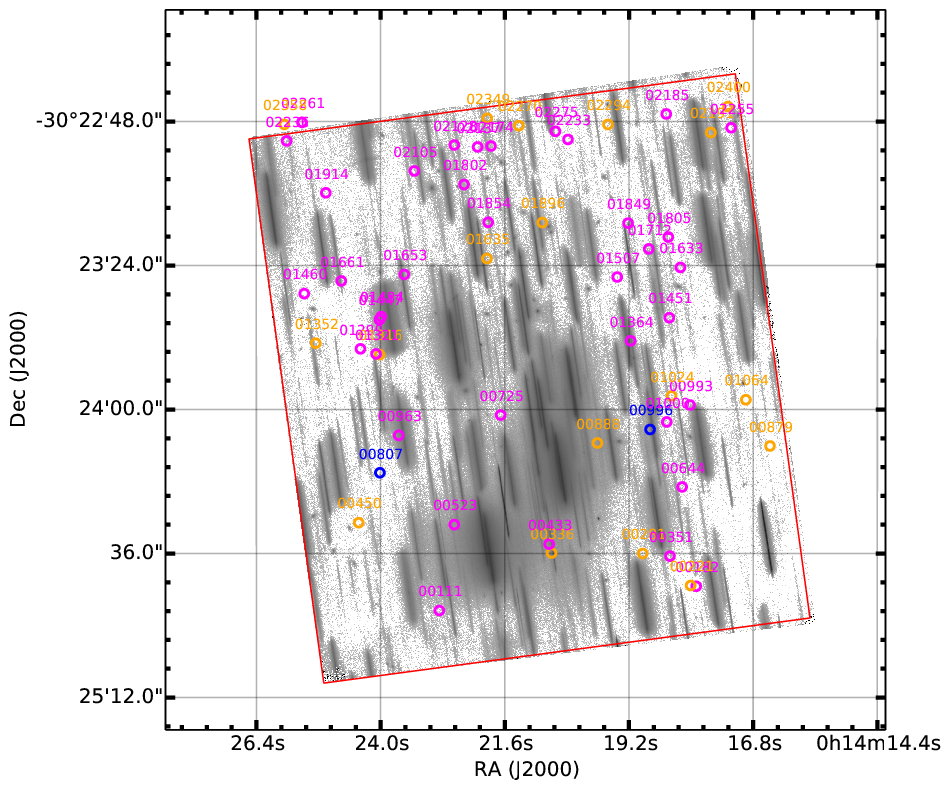}
\caption{The GLASS G102 (left) and G141 (right) grism pointings of \AB{} at two distinct P.A.s, with field-of-view shown by the 
red (P.A.=233 degrees) and green (P.A.=135 degrees) squares. The circles in all panels denote the positions of the emission line 
objects identified in this work, with color coding following the convention adopted in Figure~\ref{fig:image}.}
\label{fig:grisms}
\end{figure*}

The GLASS observations are designed to follow the 3D-HST observing strategy \citep{Brammer:2012p12977}.
The data were processed with an updated version of the 3D-HST reduction pipeline\footnote{\url{http://code.google.com/p/threedhst/}}.  Below we summarize the main steps in the reduction process of the GLASS data but refer to 
\cite{Brammer:2012p12977} and the GLASS survey paper \citep{Treu:2015p36793} for further details.

The data were taken in a 4-point dither pattern identical to the one
shown in Figure~3 of \cite{Brammer:2012p12977} to optimize rejection of bad pixels and cosmic rays
and improve sampling of the WFC3 point spread function.
At each dither position, a combination, a direct (F105W or F140W), and a grism (G102 or G141) exposure were taken.
The direct imaging is commonly taken in the filter ``matching'' the grism, i.e. pairs of F105W-G102 and F140W-G141.  However, to 
accommodate searches for supernovae and the characterization of their curves in GLASS clusters, each individual visit is designed 
to have imaging in both filters. Hence several pairs of F140W-G102 observations exist in the GLASS data. This does not affect the 
reduction and extraction of the individual GLASS spectra.

The individual exposures were turned into mosaics using AstroDrizzle from the DrizzlePac \citep{Gonzaga:2014p26307}, the  
replacement for MultiDrizzle \citep{Koekemoer:2003p31861} used in earlier versions of the 3D-HST reduction pipeline. Then all  
exposures and visits were aligned using \verb+tweakreg+, with backgrounds subtracted from the direct images by fitting a second  
order polynomial to each of the source-subtracted exposures. We subtracted the backgrounds using the master sky presented by  
\cite{Kummel:2011p33451} for the G102 grism, while for the G141 data the master backgrounds were developed by  
\cite{Brammer:2012p12977} for 3D-HST. The individual sky-subtracted exposures were combined using a pixel scale of $0\farcs06$ per  
pixel as described by \citet{Brammer:2013p27911} ($\sim$half a native WFC3 pixel) which corresponds to roughly 12\AA/pixel and  
22\AA/pixel for the G102 and G141 grism dispersions, respectively. {Figure~\ref{fig:grisms} shows} these full field-of-view 
mosaics of the two NIR grisms (G102 on the left and G141 on the right) at the two GLASS P.A. for \AB{}. The  individual spectra 
were extracted from these mosaics by predicting the position and extent of each two-dimensional spectrum according to the 
\verb+SExtractor+ \citep{Bertin:1996p12964} segmentation maps from the corresponding direct images. This is done for every single 
object and contaminations, i.e., dispersed light from neighboring objects in the direct image, so these contaminations can be 
estimated and accounted for.

\section{Identification of multiple images}
\label{sec:mul}

In this section we describe how we identify and vet multiple image
candidates using imaging (\ref{subsec:photometry}) and spectroscopic
(\ref{subsec:targeted} and \ref{subsec:blind}) data.

\subsection{Imaging data: identification and photometric redshifts}
\label{subsec:photometry}

\begin{figure*}
\begin{center}
\includegraphics[width=0.8\textwidth]{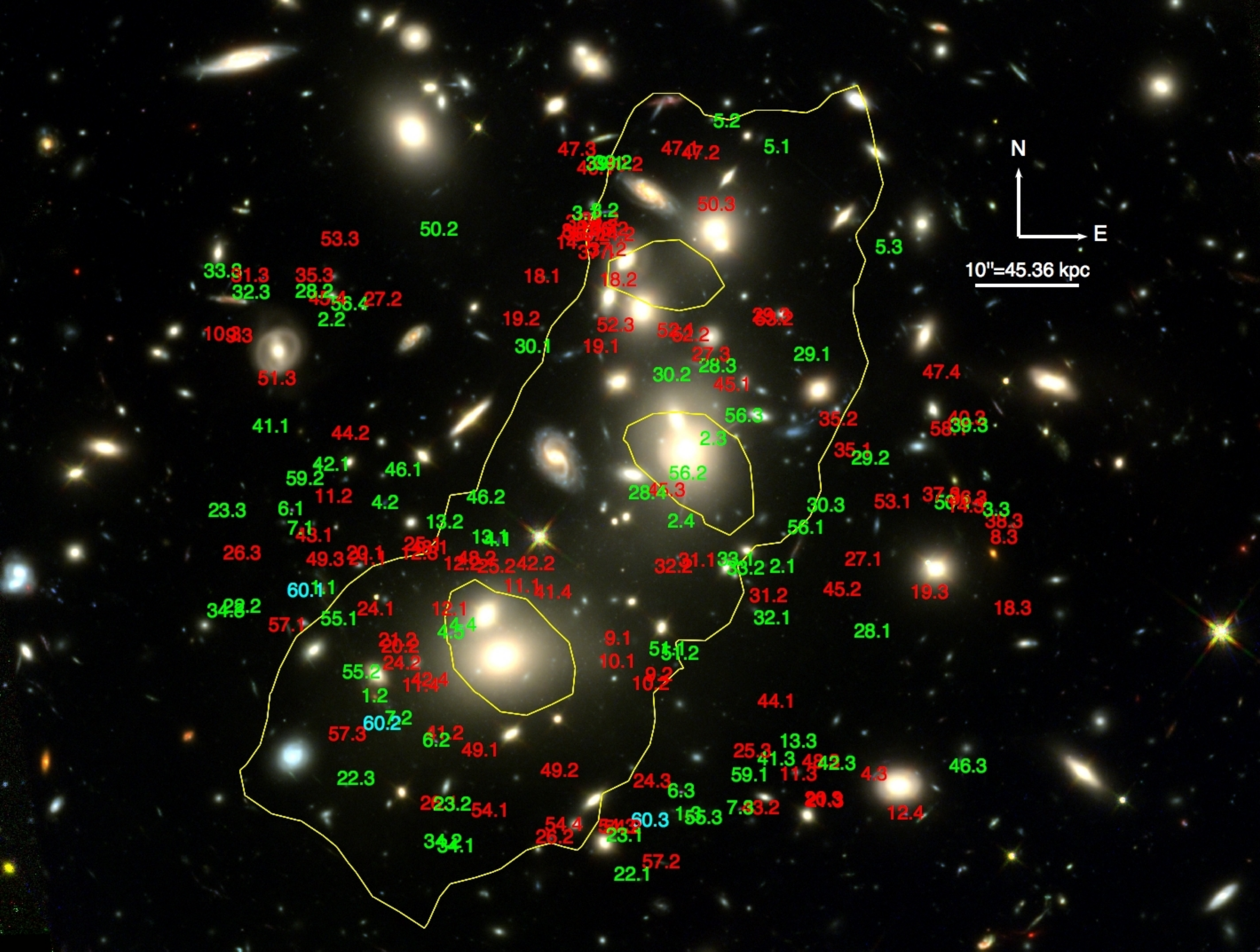}\\
\caption{All multiple images discovered to date in \AB. Green images are used in the lens model, while red images are unused for 
reasons discussed in Section~\ref{sec:mul}. Cyan images belong to a new multiple image system discovered in this work. The yellow 
line is the critical curve from our best-fit lens model at $z=9$. The RGB image is a combination of F105W, F125W and F160W.  }
\label{fig:arcs_image}
\end{center}
\end{figure*}

\citet{2011MNRAS.417..333M} published the first detailed strong lensing analysis of \AB{} identifying a total of 34 multiple 
images  (11 source galaxies) in imaging data pre-dating the HFF. With the addition of the much deeper HFF data, now a total of 
176/56  candidate arc images/systems have been identified prior to this work \citep[][see also Figure~\ref{fig:arcs_image} and 
Table~\ref{tab:mult_images}]{2014ApJ...786...60A,2015ApJ...800...18A,2014ApJ...793L..12Z,2014MNRAS.444..268R,2014arXiv1409.8663J,2014ApJ...797...98L,Ish++15}.
We identify a new multiply imaged system (i.e. system 60 in Table~\ref{tab:mult_images}) comprised of three images. 

Despite the vast number of identifications, only a handful of multiple images have been spectroscopically confirmed; arcs 4.3 and 
6.1 were spectroscopically confirmed by \citet{2014MNRAS.444..268R}, arcs 3.1, 3.2 and 4.5 by \citet{Joh++14}, and arc 18.3 by 
Cl\'{e}ment et al. (in preparation). All corresponding redshifts can be found in Table~\ref{tab:mult_images}.  When spectroscopy 
is lacking, confirming that images belong to the same source is much more difficult. Photometric redshifts alone are not adequate 
for confirmation.  \citet{Ilb++06} found that, within a given survey, the fraction of catastrophic errors in photometric redshift increases with faintness 
and redshift. Multiple images are typically faint and are necessarily at redshifts larger than the cluster redshift, which is 
$z=0.308$ for \AB{}. The mean F140W magnitude of all images in Table~\ref{tab:mult_images} is 27.11, and the mean source redshift 
is $z=2.63$. Photometry of sources in cluster fields is complicated due to blending with cluster members and the ICL.  While we do 
compute photometric redshifts in order to use the multiple images in our lens model, we do not rely on them alone to test the 
fidelity of the images.

We instead rely on color and morphology information to determine whether images belong to the same multiple image system; to first order, multiple images of the same source have identical colors. To compute colors, photometry is done using \verb+SExtractor+ in dual-image mode. We use F160W as the detection image because it detects the largest fraction of all multiple image candidates. We then measure isophotal magnitudes and errors 
in all seven photometric filters. Due to difficulties in detecting many of the multiple image candidates using the default \verb+SExtractor+ 
settings, we adopt a more aggressive set of settings for the objects with low $S/N$ and/or highly blended. We refer to the 
default \verb+SExtractor+ settings as ``cold'' mode and the more aggressive one as ``hot'' mode photometry. These are similar in spirit but not identical to those adopted by \citet{Guo+13}. Even with the ``hot'' mode settings, we cannot detect all of the multiple image candidates, though the detected fraction is vastly increased over the ``cold'' mode settings. 
Using the seven HFF photometric filters, we compute 4 colors for each image: 
$F435W-F606W$, $F814W-F105W$, $F125W-F140W$, and $F140W-F160W$\footnote{Note that the last two colors are not independent due to 
the repetition of $F140W$. We chose to repeat one filter to increase the number of color bins.}. The colors are computed 
within a fixed aperture (MAG\_APER) that is $0.4''$ in diameter. We compute a reduced ``color-$\chi^2$'' for each image:
\begin{equation}
  \chi^2_{I \nu}  = {1 \over N-1} \sum_{i=1}^{N} \left({C_i - \bar{C}_i \over \sigma_i} \right)^2,
\end{equation}
where $i$ runs over the number of colors, $N$ is the total number of colors we are able to measure, $C_i$ is the $i$-th color, 
$\bar{C_i}$ is the inverse variance-weighted mean color of all images in the system, and $\sigma_i$ is the uncertainty in the 
$i$-th color. 

Multiple images of the same source also have predictable morphologies. In rare cases, more than one images of the same source 
possess a number of uniquely identifiable features. For instance, there are two such systems in \AB, i.e., systems 1 and 2. The 
counterparts to systems 1 and 2 are systems 55 and 56, respectively. We include both counterpart systems in the lens model at the 
spectroscopic redshifts of systems 1 and 2 that we measure in this work. All multiple images in Table~\ref{tab:mult_images} are 
visually inspected. We assign each image a grade that determines the likelihood that it is part of the system to which it is 
assigned. {We perform this grading exercise in a lens model-independent fashion; we do not make any assumptions about the location of the critical curves relative to the graded images. There are, however, configurations of multiple images that are impossible to achieve through gravitational lensing by galaxy clusters, such as three individual images (not part of an elongated arc) on the same side of, and very distant from the cluster core with no counter-images. Other information such as surface brightness and symmetry can be incorporated independently of an assumed mass distribution, and we rely on this information much more heavily in assigning the morphology grade.} The grading scheme 
based on morphological similarity is as follows:

\begin{description}
    \item[4] Image is definitely part of the system
	\item[3] Image is very likely part of the system
	\item[2] Image is potentially part of the system
	\item[1] Image is very unlikely part of the system
	\item[0] Image is definitely not part of the system
\end{description}

\noindent
Two inspectors (A.H. and M.B.) independently assign a grade to each image. {The inspectors use several RGB images of the full HFF depth that span the full HST spectral coverage to assign the grade for each image.} The two grades are then summed to get the reported morphology grade. Examples of multiple images 
that receive high and low morphology grades are shown in the Appendix.

We use the color and morphology information together to determine whether to include a multiple image in our model. The joint criteria are:
\begin{equation}
  (\chi^2_{I \nu} < 1.5 \quad \lor \quad \textrm{M} > 5) \quad \land \quad M>0,
\end{equation}
where M is the summed morphology grade from each inspector, which ranges from 0-8. In cases where the contamination by foreground 
objects, clusters members or ICL is severe, we rely only on the morphology criterion, $M>5$. The particular $\chi^2_{I \nu}$ 
threshold value of $1.5$ was chosen because it is in the typical range for a good reduced chi-square test, and most of the images 
in spectroscopically confirmed systems are below this value. $M>5$ is chosen because in the least confident case that obeys this, 
$M=6$, the modelers either both think the image is very likely part of the system ($M=3$) or one thinks the image is potentially 
part of the system ($M=2$), while the other is sure of it ($M=4$). 

\begin{figure*}
\begin{center}
    \includegraphics[width=0.8\textwidth]{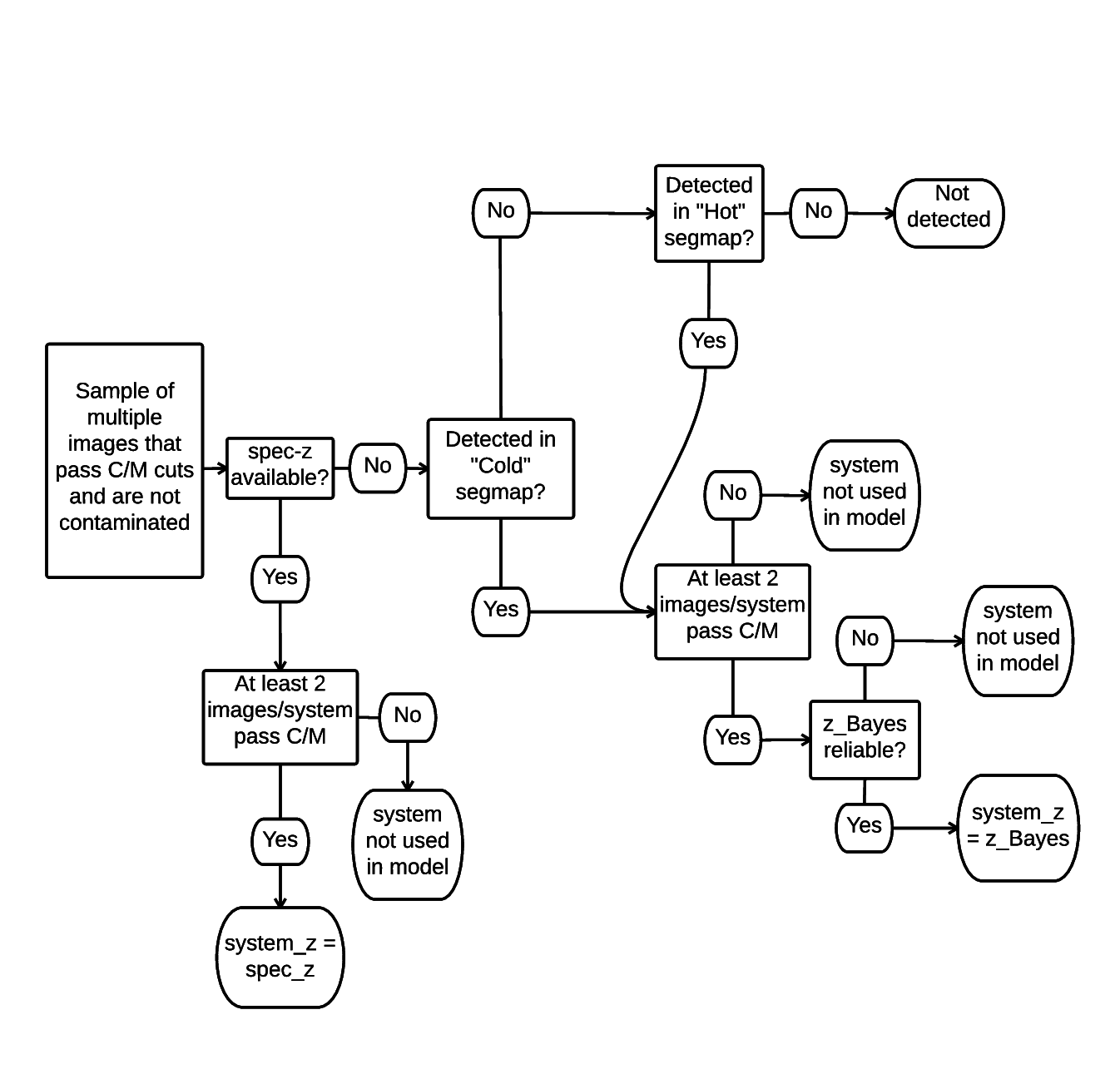}\\
    \caption{Flow chart describing our procedure for assigning photometric redshifts to multiple image systems. Two segmentation 
        maps, a ``hot'' and a ``cold'' version, were used for source detection. The detection and deblending thresholds are set to 
        the SExtractor defaults in the cold segmentation map. Objects detected in the cold segmentation map, typically the 
        brighter and more isolated ones, have more accurate photometry. The hot segmentation map was created using extremely 
        aggressive detection and deblending thresholds. It detects the majority of the remaining multiple images that are not 
        detected in the cold version.  $\textrm{C}/\textrm{M}$ cuts refer to the color/morphology cuts used to purify the sample 
        of multiple images.  $z_{\mathrm{Bayes}}$ refers to the redshift obtained from combining multiple redshifts in a 
    hierarchical Bayesian model \citep{Dahlen:2013p33380}. $z_{\mathrm{Bayes}}$ is considered reliable if it is larger than the 
redshift of \AB, $z=0.308$, and it is not multi-valued. See the Appendix for examples of multiple image systems that pass and fail 
some of the tests in this flow chart. }
\label{fig:flow_chart}
\end{center}
\end{figure*}

For a multiple image system to constrain the lens model, we must
estimate its redshift. Having photometric redshift measurements for
multiple images of the same source can provide a tighter constraint
than a single measurement. Individual photometric redshifts are
computed using EAZY \citep{Brammer:2008p13280}. We then use a
hierarchical Bayesian model to obtain a single redshift probability
density function for each system \citep{Dahlen:2013p33380}. The mode
of this probability density function will be referred to as
$z_{\textrm{Bayes}}$. Only non-contaminated objects contribute to
calculating $z_{\textrm{Bayes}}$. We graphically outline the procedure
for measuring and including photometric redshifts as inputs to our
lens model in Figure~\ref{fig:flow_chart}. 2/57 systems (36 and 52) are
entirely contaminated, so we do not compute $z_{\textrm{Bayes}}$ for
those systems. 14/57 systems have $z_{\textrm{Bayes}} <
z_{\textrm{Cluster}}$, and thereby are not included in lens modeling (a
fraction of those have poorly constrained posteriors, monotonically
declining from 0; they are highly uncertain and considered unreliable;
we label them by assigning $z_{\rm Bayes}=0.01$). 5/57 systems have a
multi-modal or extremely broad Bayesian redshift distribution. We similarly do
not include these systems in the lens modeling. For systems where only one image passed the
color/morphology criteria and that image is not contaminated, we report
$z_{\textrm{Bayes}}$ in Table~\ref{tab:mult_images}, but we do not
include these systems in the modeling. We show the posterior redshift
distributions for some of these cases in the Appendix. System 18
consists of a spectroscopically confirmed image (18.3), but the system as
a whole does not pass all criteria required to be considered a
multiple image system. Our screening rules out all the above-mentioned
systems and delivers a secure set of 25/57 multiple arc systems that
are used in lens modeling.

\subsection{Targeted GLASS spectroscopy}
\label{subsec:targeted}

\begin{figure*}
  \centering
  \includegraphics[width=\textwidth]{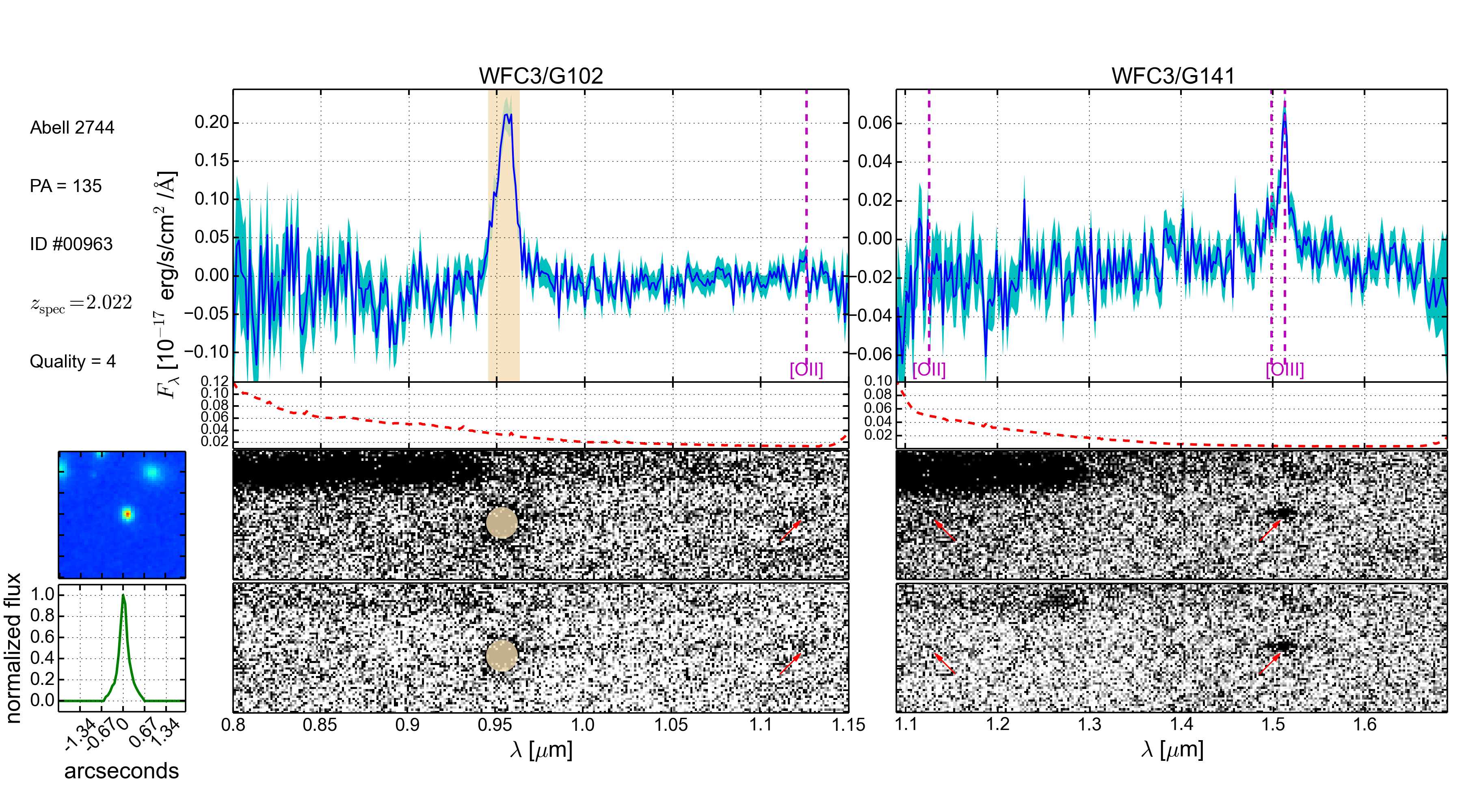}\\
  \includegraphics[width=\textwidth]{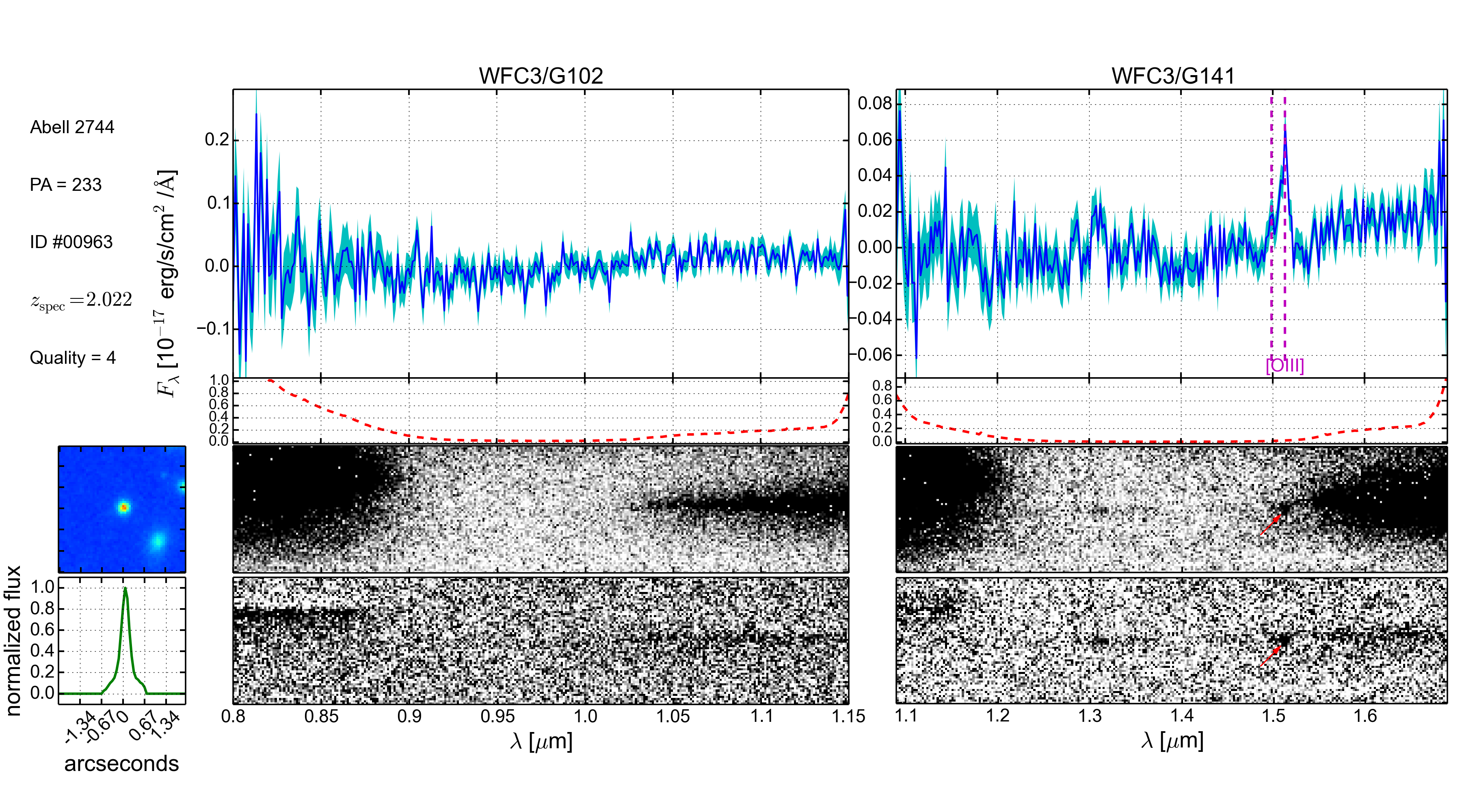}
  \caption{Emission line detection results on object ID \#00963 (arc 6.1) at the two P.A.s displayed in the two sub-figures 
  accordingly. In each sub-figure, the two panels on the first row show the observed 1-dimensional spectra, where the 
  contamination subtracted flux is denoted by the blue solid line and the noise level by the cyan shaded region. The two panels on 
  the second row give the corresponding contamination model in red dashed line. For the four panels directly underneath, the top 
  two display the interlaced 2-dimensional spectra whereas the bottom two have contamination subtracted. In the 1- and 
  2-dimensional spectra, the identified emission lines are denoted by vertical dashed lines in magenta and arrows in red, 
  respectively. The wheat colored regions cover contamination model defects. The two panels on the far left refer to the 
  2-dimensional postage stamp created from the HFF co-adds through drizzling (on top) and the 1-dimensional collapsed image (on 
  bottom). Note that these two panels share the same x-axis along the grism dispersion direction. Some ancillary information can 
  also be seen in the upper left corner in each sub-figure.}
  \label{fig:ELarc6.1}
\end{figure*}

\begin{figure*}
  \centering
  \includegraphics[width=\textwidth]{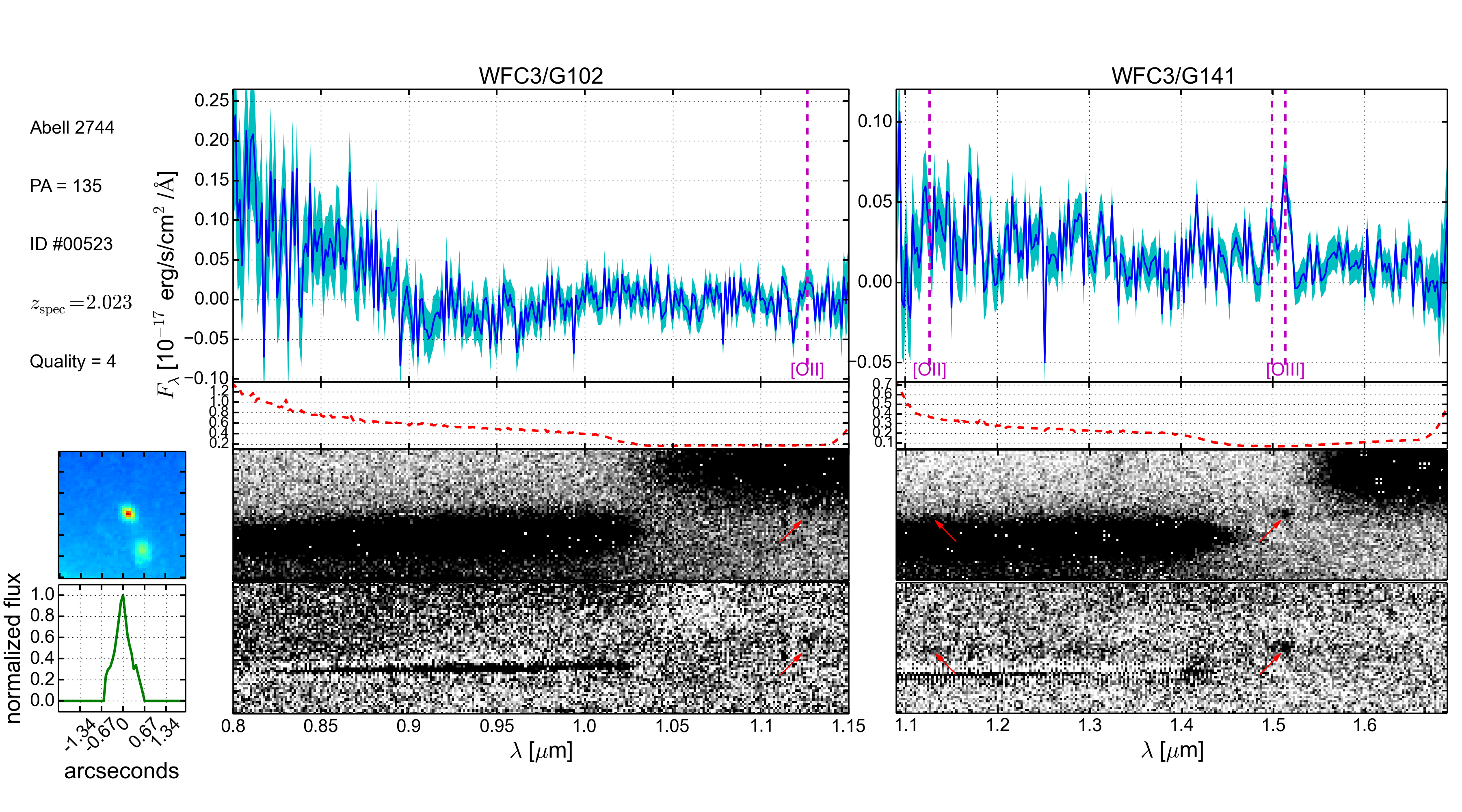}\\
  \includegraphics[width=\textwidth]{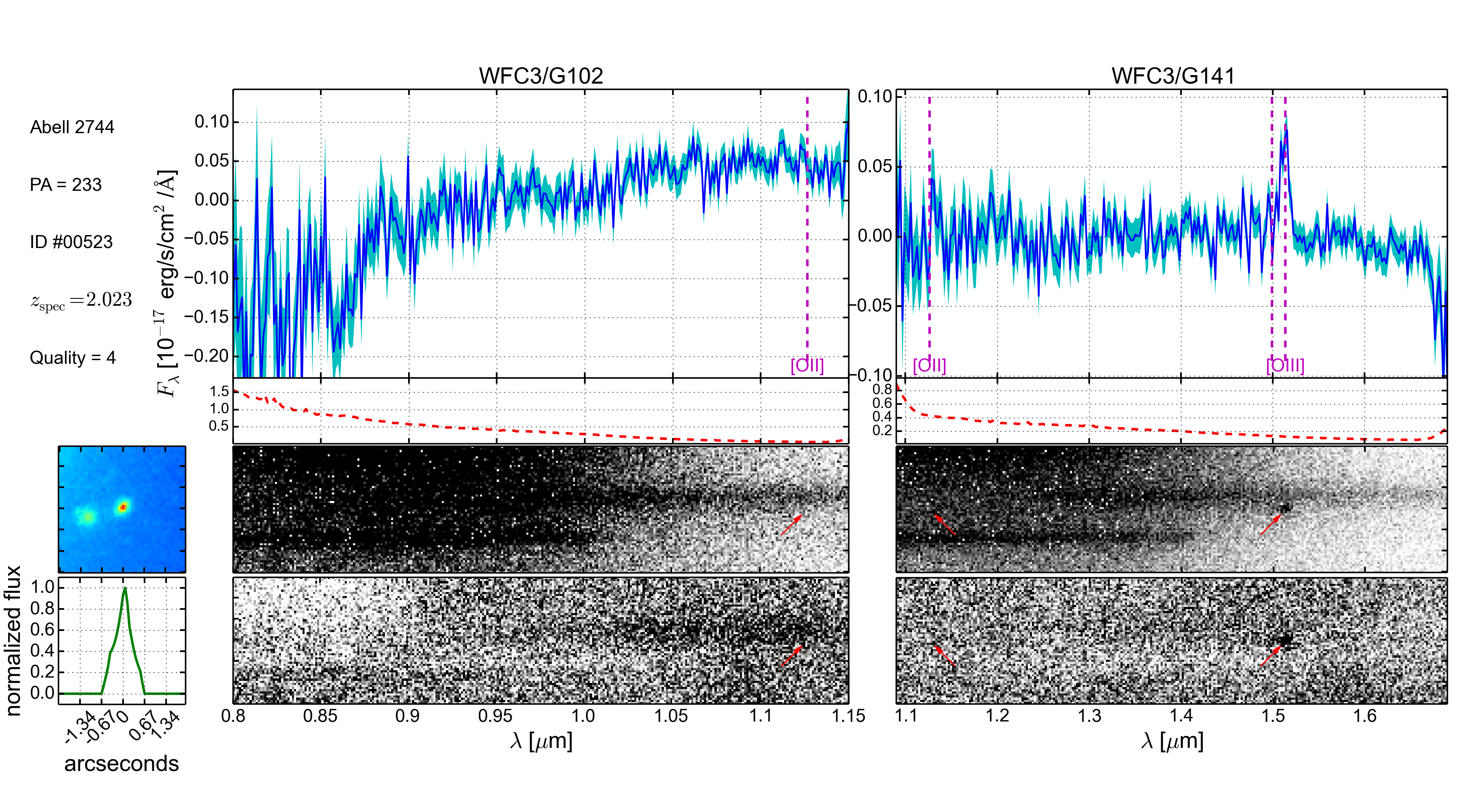}
  \caption{Same as Figure~\ref{fig:ELarc6.1}, except that object ID \#00523 (arc 6.2) is shown.}
  \label{fig:ELarc6.2}
\end{figure*}

\begin{figure*}
  \centering
  \includegraphics[width=\textwidth]{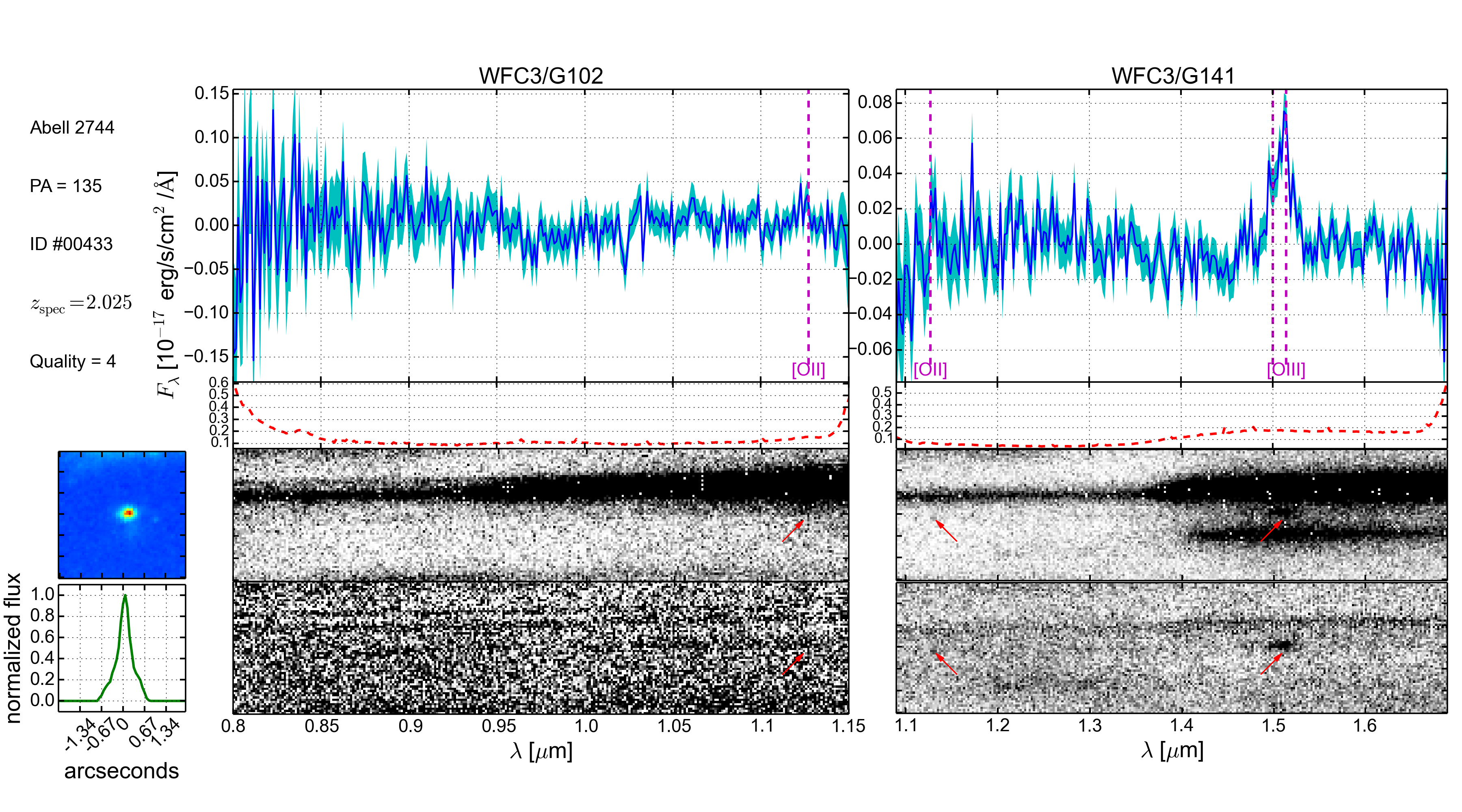}\\
  \includegraphics[width=\textwidth]{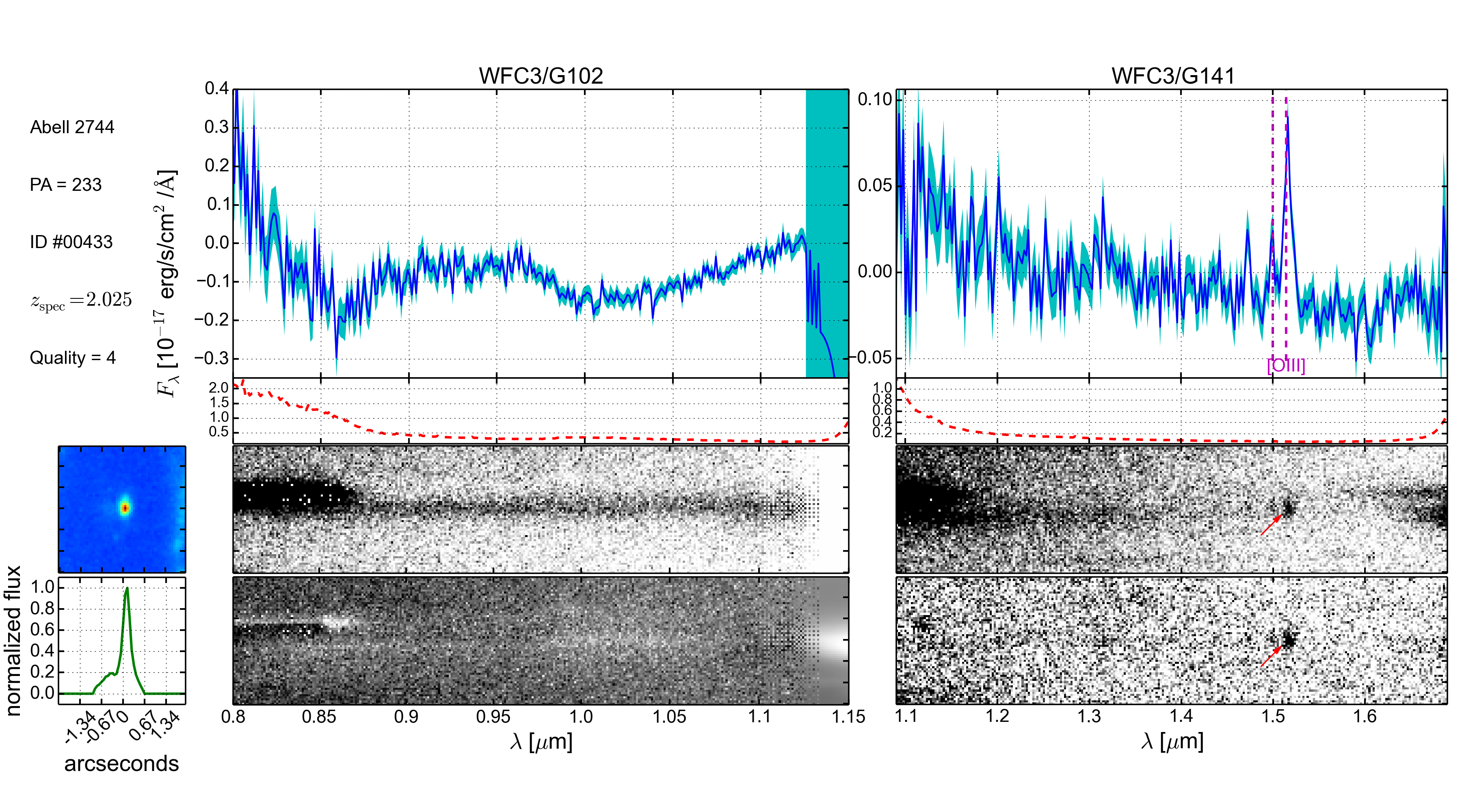}
  \caption{Same as Figure~\ref{fig:ELarc6.1}, except that object ID \#00433 (arc 6.3) is shown. Note here WFC3/G102 at the second P.A.  
  is cut off on the right due to grism defect.}
  \label{fig:ELarc6.3}
\end{figure*}

\begin{figure*}
  \centering
  \includegraphics[width=\textwidth]{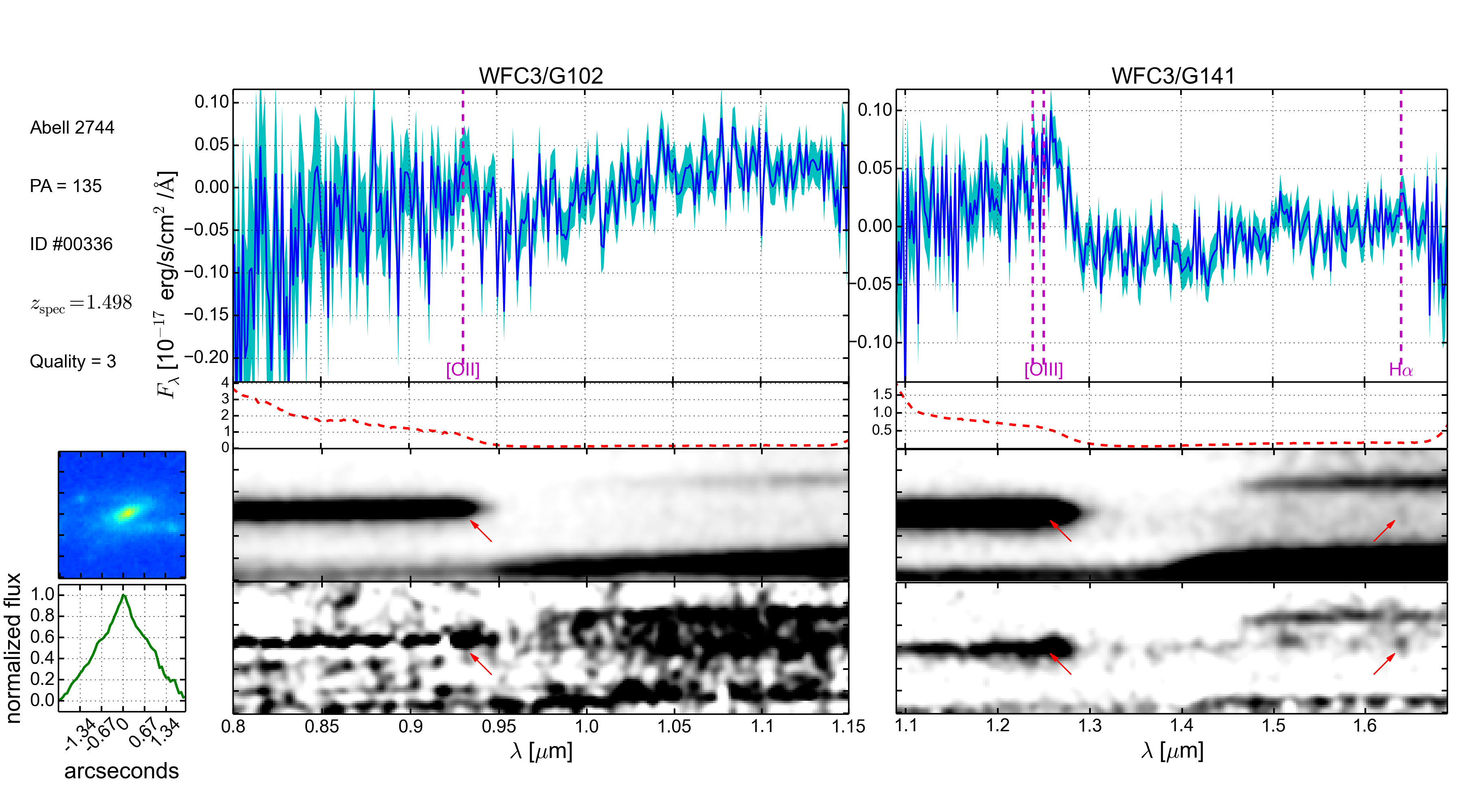}\\
  \includegraphics[width=\textwidth]{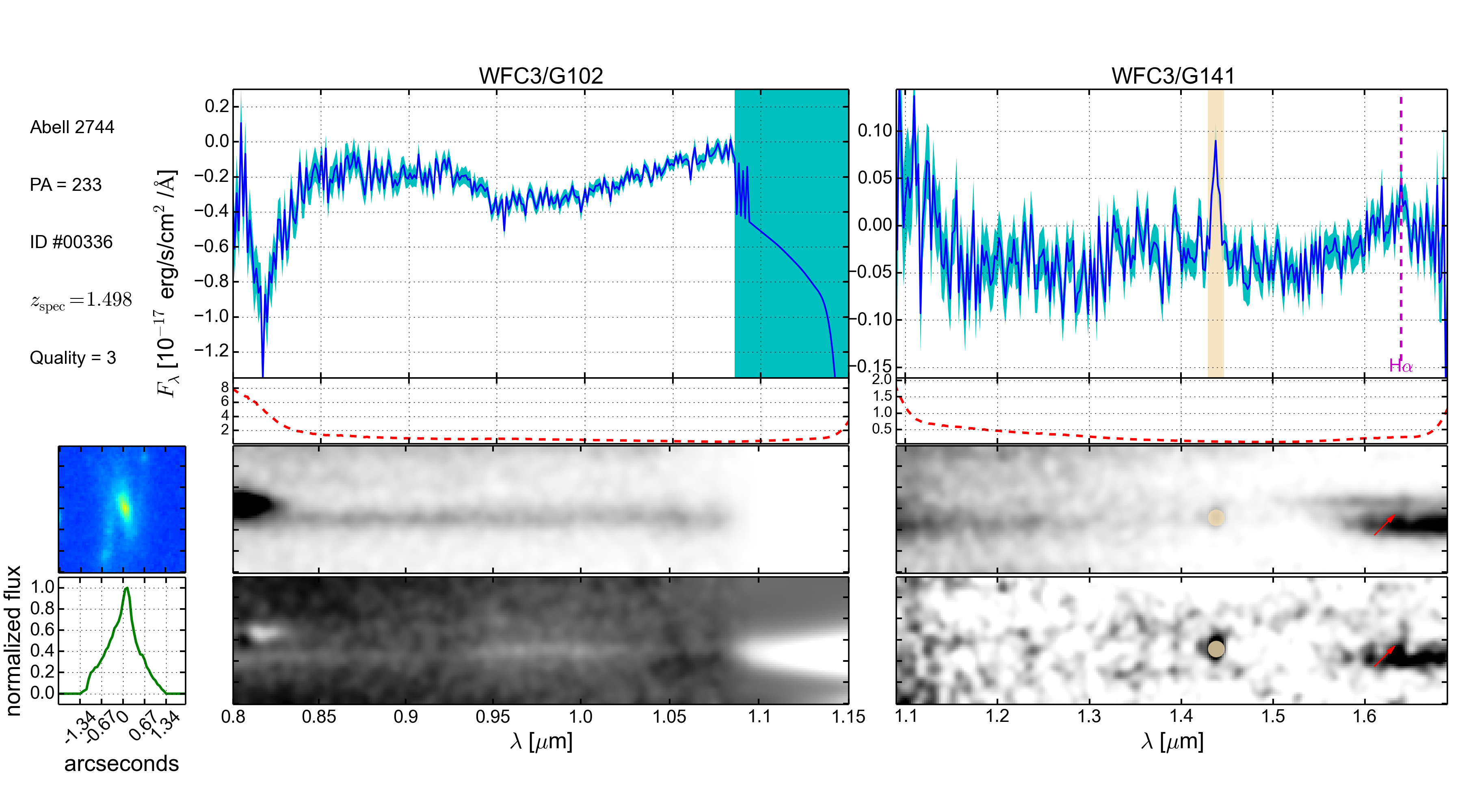}
  \caption{Same as Figure~\ref{fig:ELarc6.1}, except that object ID \#00336 (arc 1.3) is shown. Moreover, the 2-dimensional
  spectra are smoothed, whilst the 1-dimensional spectral resolution remains unchanged.}
  \label{fig:ELarc1.3}
\end{figure*}

\begin{figure*}
  \centering
  \includegraphics[width=\textwidth]{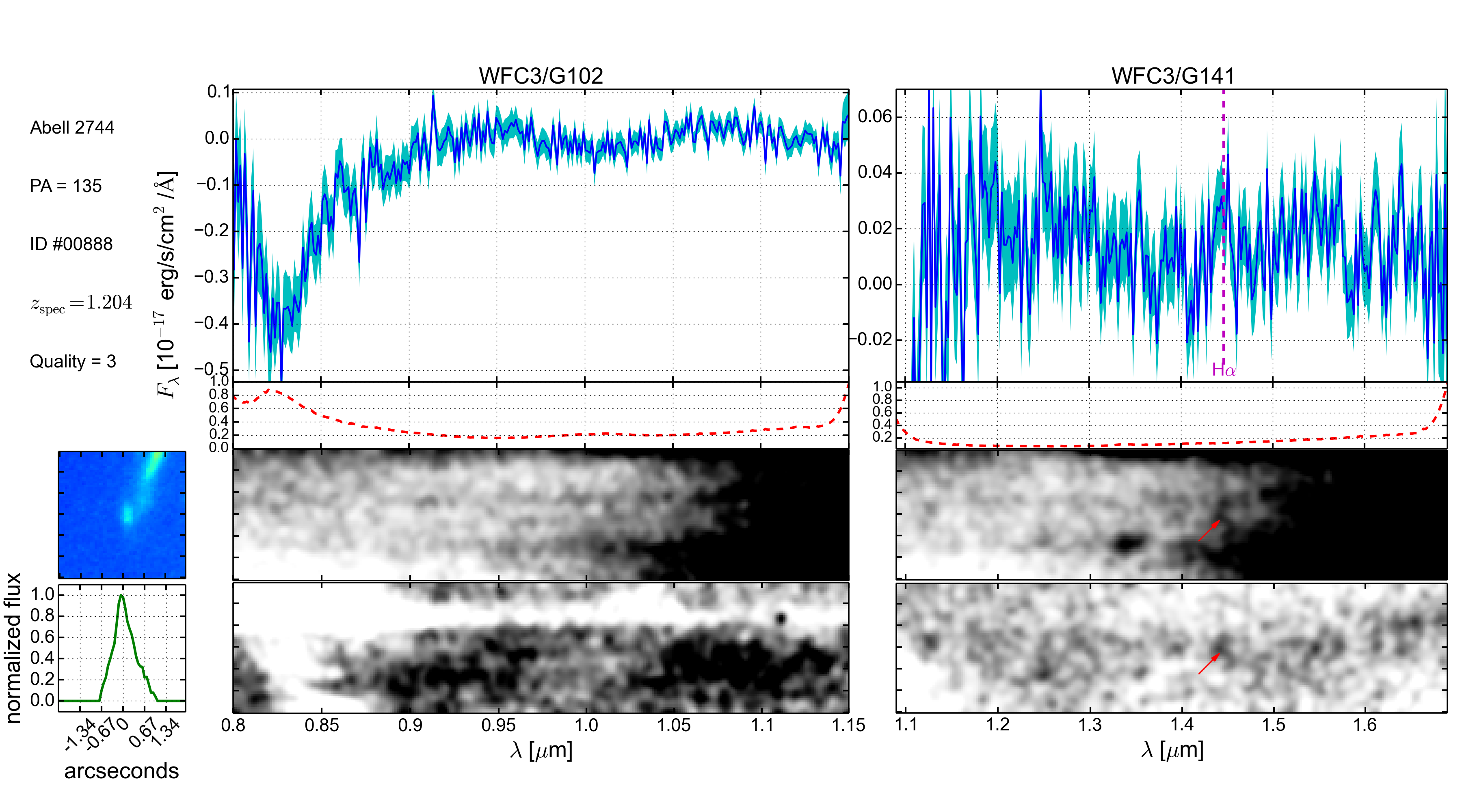}\\
  \includegraphics[width=\textwidth]{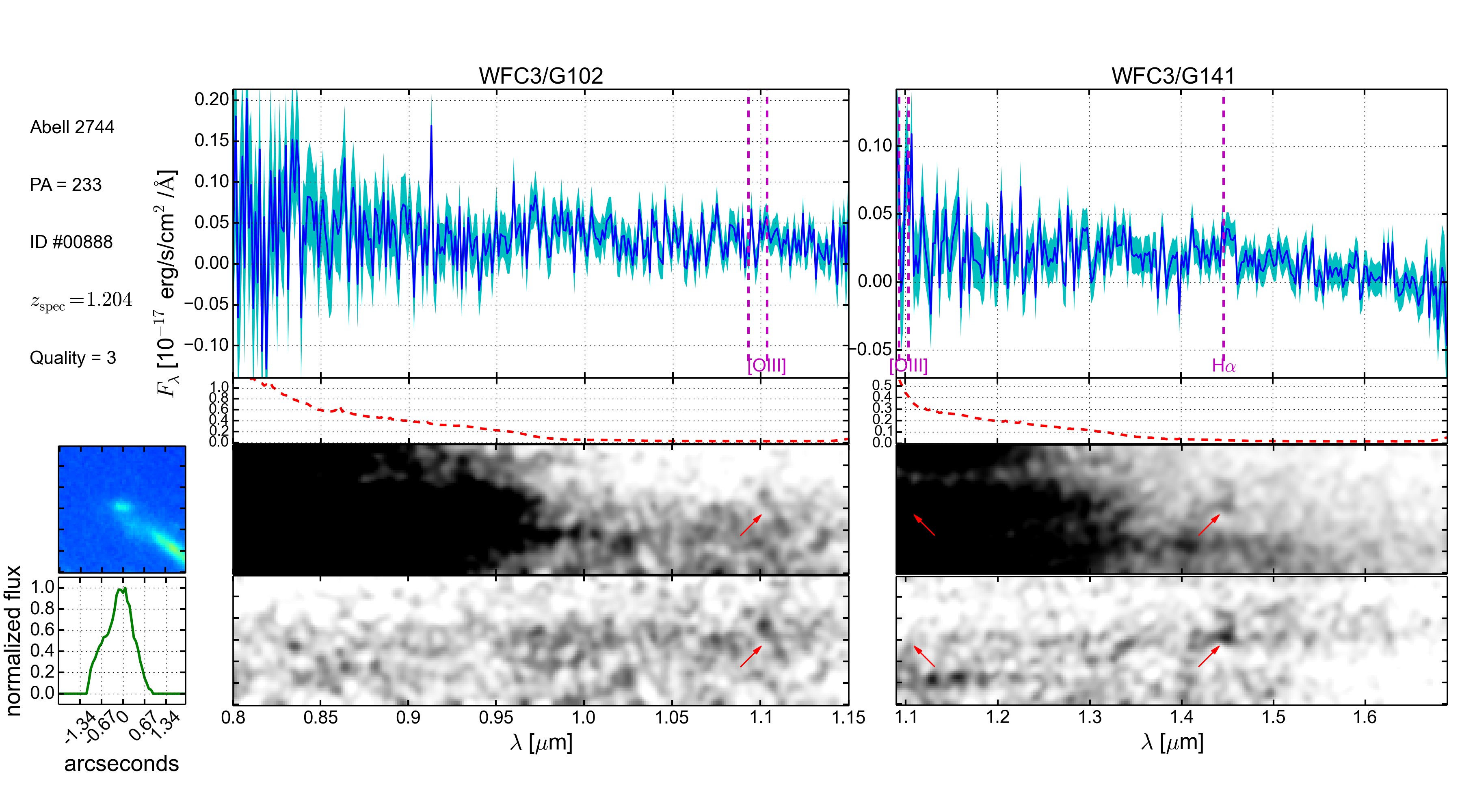}
  \caption{Same as Figure~\ref{fig:ELarc1.3}, except that object ID \#00888 (arc 56.1) is shown.}
  \label{fig:ELarc56.1}
\end{figure*}

GLASS spectroscopy was carefully examined for a total of \NimgTOT{}
multiply lensed arc candidates mostly seen on both P.A.s with the goal
of measuring spectroscopic redshifts.  As described in GLASS paper I,
each spectrum was visually inspected by multiple investigators
(X.W. and K.B.S.) using the GLASS Graphic User Interfaces (GUIs) GLASS
Inspection GUI (GiG) and GLASS Inspection GUI for redshifts (GiGz) The
results were then combined and a preliminary list of arcs with
emission lines was drafted. In the end, another round of double-check
by re-running GiGz was also executed to make sure no potential
emission lines were missed. Following GLASS procedure, a quality flag
was given to the redshift measurement: Q=4 is secure; Q=3 is probable;
Q=2 is possible; Q=1 is likely an artifact. As described in paper I,
these quality criteria take into account the signal to noise ratio of
the detection, the probability that the line is a contaminant, and the
identification of the feature with a specific emission line. For
example, Q=4 is given for spectra where multiple emission have been
robustly detected; Q=3 is given for spectra where either a single
strong emission line is robustly detected and the redshift
identification is supported by the photometric redshift, or when more
than one feature is marginally detected; Q=2 is given for a single
line detection of marginal quality. As shown in Table~\ref{tab:ELtot},
new spectroscopic redshifts were obtained for \NimgELtot{} images in
total, corresponding to \NsysELtot{} systems. Among them, \NimgELhiQ{}
high-confidence (with quality flags 3 or 4) spectroscopic redshifts
were measured for arcs 1.3, 6.1, 6.2, 6.3, 56.1. The spectra of these
objects are shown in
Figures~\ref{fig:ELarc6.1}--\ref{fig:ELarc56.1}. In particular for arc
6.1, our spectroscopic redshift matches that reported by
\citet{2014MNRAS.444..268R}, and we provide the first spectroscopic
confirmation that 6.2 and 6.3 are images of the same system.
{We note that our measured redshift for arc 
56.1 $z_{\textrm{spec}}=1.20$ (Q=3; probable) differs signficantly
from that given by \citet{Johnson:2014p37801} for arc 2.1
$z_{\textrm{spec}}=2.2$ (possible), even though the two systems are
likely to be physically connected.  Our measurement is based on three
pieces of evidence. First, we detect a spectral feature in G141 at
both P.A.s (see Figure~\ref{fig:ELarc56.1} for details) with
sufficiently high signal to noise ratio to study its spectra
shape. The feature is better described by a single line (identified by
us as H$\alpha$ at $z=1.2$) rather than a doublet like [OIII] ($\Delta
\chi^2=2.4$). Second, a line is marginally detected in one of the G102 spectra at exactly the wavelength expected for [OIII] $z=1.2$.
Third, the wide spectral coverage of our data and the data available
in the literature rule out the possibility of the feature we see in
G141 being other prominent lines such as MgII and [OII]. Taking into
account both the evidence and previous results, we assign a quality
flag of Q=3 (probable).}

The uncertainty on our spectroscopic redshift measurements is limited
by the resolution of approximately 50\AA\ and by uncertainties in the
zero point of the wavelength calibration. By comparing multiple
observations of the same object we estimate the uncertainty of our
measurements to be in the order of $\Delta z\sim 0.01$. This is
sufficient for our purposes and we will not pursue more aggressive
approaches to improve the overall redshift precision
\citep[e.g.][]{Brammer:2012p12977}.

\subsection{Blind search in GLASS data}
\label{subsec:blind}

The targeted spectroscopy done in Section~\ref{subsec:targeted} could
potentially miss some multiply imaged sources that are not identified
photometrically. In order to increase the redshift completeness of
emission line sources (both multiply and singly imaged), we also
conducted a blind search within the entire grism field-of-view. As a
first step, one of us (T.T.)  visually inspected all the 2D grism
spectroscopic data, the contamination models, and residuals after
contamination for each of the 2445 objects in the prime filed of \AB{}
given by the GLASS catalog.  This yielded a list of 133 candidate
emission line systems that were later on inspected using the GLASS
GUIs GIG and GIGz by two of us (X.W.  and K.B.S.) to confirm emission
line identifications and measure redshifts.  In order to search for
previously undiscovered multiple images we inspected each set of
objects with mutually consistent redshift.  None of the sets of
galaxies at the same spectroscopic redshifts are consistent with being
multiply lensed images of the same source. Some of them are ruled out
because of their position in the sky, while others are ruled out
because their colors and morphology are inconsistent with the lensing
hypothesis.

\begin{figure}[ht]
    \includegraphics[width=0.5\textwidth]{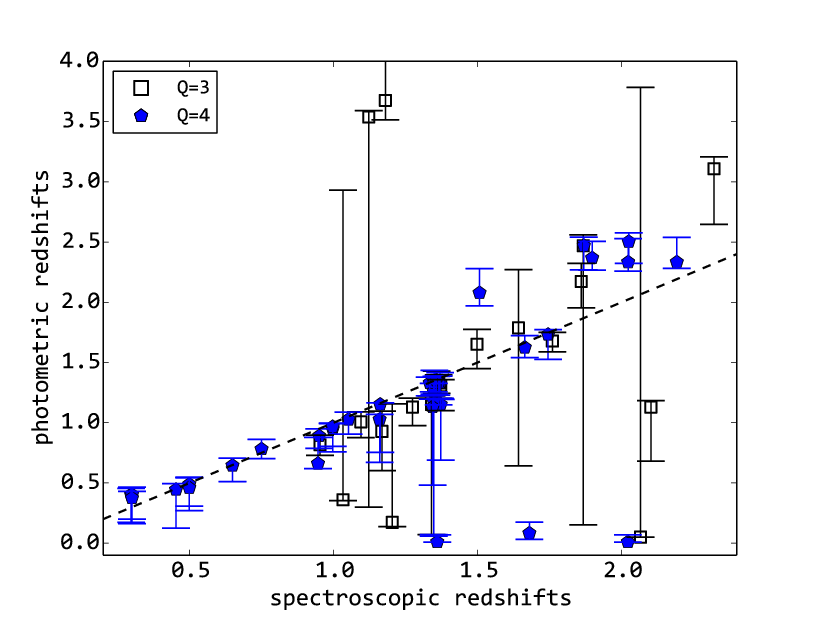}
\figcaption{
Comparison between the spectroscopic and photometric redshifts for the
55 objects with high-confidence emission lines (quality flags 3 or 4).
We also show the 1-$\sigma$ error bars (enclosing 68\% of the total
probability) around the photometric redshifts. There is reasonably
good agreement between photometric and spectroscopic redshifts, with
25 out of 55 spectroscopic redshifts within the photometric redshift
error bars. This is acceptable considering that the photometric redshift
uncertainties only include the random component, even though they
suggest that an additional systematic uncertainty component is
present. In order to account for this systematic uncertainty, for the
systems used to build the lens model we added 20\% in quadrature to
our photometric errors.
\label{fig:ELobjs_photz}}
\end{figure}

Nonetheless, we compiled a list of singly imaged emission line
objects, consisting of \NELQtwo{}, \NELQthree{}, and \NELQfour{}
quality 2, 3, and 4 spectroscopic redshift measurements respectively,
which are color coded in Figure~\ref{fig:image}. Among them, the
high-confidence (with quality flag 3 or 4; orange and magenta circles
in Figure~\ref{fig:image}) emission line identifications are also
included in Table~\ref{tab:ELtot}. As mentioned in
Section~\ref{subsec:photometry}, via running EAZY on the full-depth
seven-filter HFF imaging data, we were able to measure photometric
redshifts for those objects as well. As a result, a comparison between
spectroscopic and photometric redshift measurements is possible, as
displayed by Figure~\ref{fig:ELobjs_photz}. We find that 25/55
photometric redshifts agree within their 1-$\sigma$ uncertainties with
corresponding spectroscopic redshifts, when nebular emission lines are
included in the fitting template. This suggests the presence of
additional systematic errors that are likely related to the
photometric redshift fitting method. In order to account for the
unknown systematics, we increase the photometric redshift
uncertainties for the sources used in the construction of the lens
model.

We double-checked our spectroscopic redshift measurements by
re-running GiGz on all the objects and also cross-checked the photometric
redshift measurements through re-fitting the photometric redshifts
using a different method by a subset of the authors (R.A., M.C., E.M)
without knowing previous results. The general conclusions about the
agreement between spectroscopic and photometric analyses remains
unchanged.

\section{Gravitational Lens Model}
\label{sec:mass}

\begin{figure*}[ht]
\plottwo{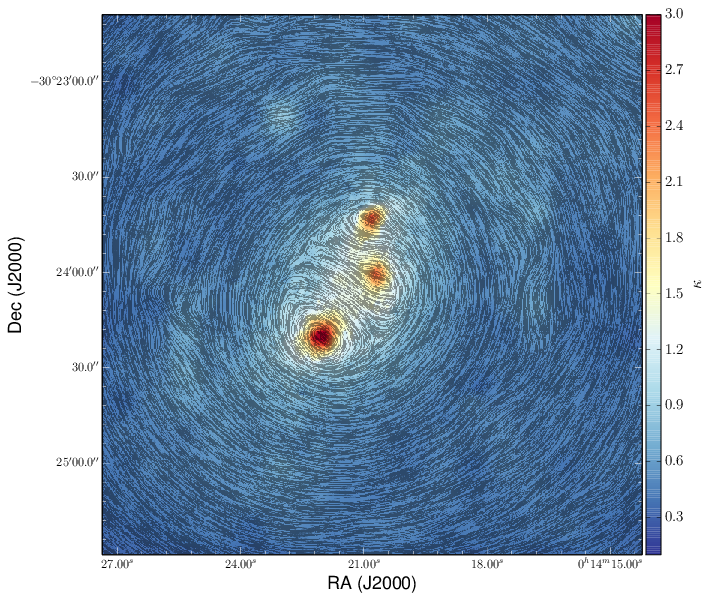}{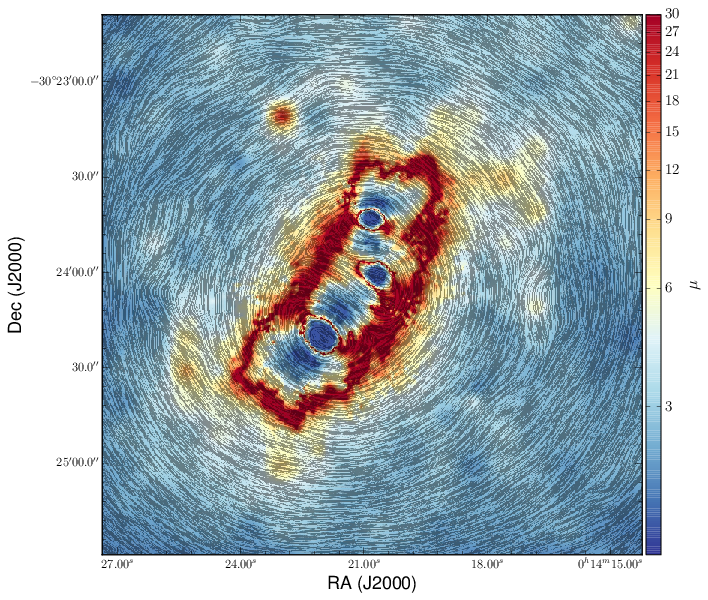}
\figcaption{Convergence (left), $\kappa$ and flux magnification (right), $\mu$, maps of \AB{} produced by our lens model for a source at $z=9$. Maps cover $3.5 
  \times 3.5 \, \textrm{arcmin}^2$. In both maps, the gray stick pattern indicates the phase angle of the shear. The figures are made using software written by Dan Coe.
\label{fig:lens_maps}}
\end{figure*}

Our lens modeling method, SWUnited (\citealp{bradac09},
\citealp{bradac05}), constrains the gravitational potential within a
galaxy cluster field via $\chi^2$ minimization. It takes as input a
simple initial model for the potential. A $\chi^2$ is then calculated
from strong and weak gravitational lensing data on an adaptive,
pixelated grid over the potential established by the initial
model. The number of grid points is increased and the $\chi^2$ is
recalculated. Once the minimum is found, and convergence is achieved,
derivative lensing quantity maps, such as convergence ($\kappa$),
shear ($\gamma$) and magnification ($\mu$), are produced from the
best-fit potential map. Errors in these quantities are obtained via the method described below.
Maps of the convergence and magnification
are shown in Figure~\ref{fig:lens_maps}.

A previous model of \AB{} using pre-HFF data was created using the same lens modeling code. The model was created as part of a 
call by STScI to model the HFF clusters, and it appears on the publicly accessible HFF lens modeling website as the Bradac v1 
models\footnote{\url{http://www.stsci.edu/hst/campaigns/frontier-fields/Lensing-Models}}. The previous model was constrained using 
44 total multiple images belonging to 11 distinct systems. The weak lensing constraints were obtained by one of the modelers, 
Julian Merten, and distributed to all participating modeling teams. The same weak lensing constraints were used in the model that 
appears in this work. This model is also made available to the public on the HFF lens modeling website as the Bradac v2 model.  In 
the v1 model, magnification uncertainties were estimated by bootstrap-resampling the weak lensing galaxies. In this work, however, 
we took a different approach to estimate uncertainties, one that we expect more accurately represents the true uncertainties. 
Because the number of multiple image systems used in this model is much larger than in the v1 model, 72 total multiple images 
belonging to 25 distinct systems, we bootstrap-resampled the multiple image systems that were not spectroscopically confirmed. 
These are the systems for which we use $z_{\textrm{Bayes}}$ in the lens model. {We assess the impact of photometric 
redshift uncertainty on the derived lensing quantities by resampling the redshift of each system lacking spectroscopic 
confirmation from their full $z_{\textrm{Bayes}}$ posteriors\footnote{We exclude values of the redshift 
$z<z_{\textrm{cluster}}+0.1$ when resampling from the $z_{\textrm{Bayes}}$ posteriors because they are unphysical.}. We compare the 
variance in magnification due to redshift uncertainty with the variance in magnification due to bootstrap-resampling the multiple 
image systems, finding that the latter is dominant. We nonetheless propagate both sources of error when reporting the errors on 
all derived lensing quantities in this work.}

As a test of the improvement of the lens model with the addition of the new multiple image constraints from the HFF data, we 
calculate the magnification of SN HFF14Tom, a Type Ia Supernova (SN Ia) at $z=1.35$ discovered in the primary cluster field of 
\AB{} (\citealp{Rod++15}). We compare the magnification predicted by our lens model with the magnification calculated directly 
from a comparison with other SNe Ia at similar redshifts, $\mu=2.03\pm0.29$ (\citealp{Rod++15}). The magnification predicted by 
the v1 model, using pre-HFF data was $\mu_{\mathrm{best}} = 3.15$, with $\mu_{\mathrm{median}} = 2.45^{+0.19}_{-0.16}$ ($68\%$ 
confidence).  {The new model presented in this work, v2, predicts a consistent magnification of $\mu_{\mathrm{best}} = 
2.23$, with $\mu_{\mathrm{median}} = 2.24^{+0.07}_{-0.08}$ ($68\%$ confidence).} The improved lensing constraints significantly 
improve the accuracy as well as the precision according to this test. We note that while we were not blind to the magnification of 
the supernova predicted by \citet{Rod++15} when producing the v2 lens model, we did not use the magnification as an input to our 
model. 

\subsection{Comparison with previous work}
\label{sec:compare}

\begin{figure}
    \centering
    \includegraphics[width=0.5\textwidth]{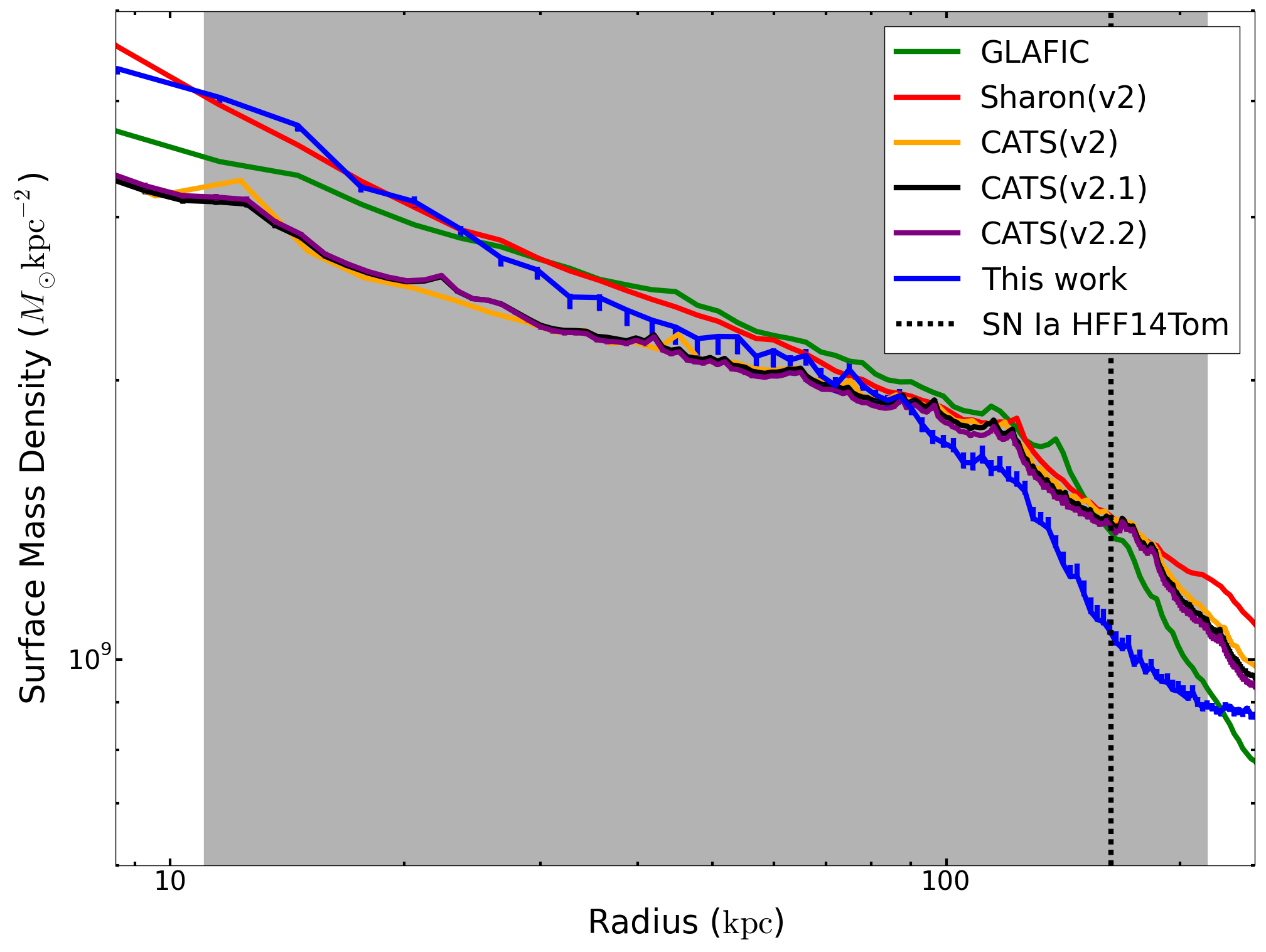}
    \caption{Surface mass density profile for the lens model obtained in this work compared to several recently published lens 
        models of \AB. The Sharon v2 model is presented by \citet{Joh++14}, the GLAFIC model by \citet{Ish++15}, the CATS(v2) 
        model in \citet{2014arXiv1409.8663J}, and the CATS(v2.1) and CATS(v2.2) models by Jauzac et al. (2015, private 
        communication).  The shaded gray region indicates the radii over which multiple image constraints are available. The 
        models agree best within this region, and they begin to significantly disagree at radii $\gtrsim 200$kpc. The radius is 
        measured from the center of the BCG. Error bars shown for our model represent $68\%$ confidence. Gaussian 1-$\sigma$ error 
        bars are included on all three CATS models, but are almost entirely too small to discern.
\label{fig:surfcompare}}
\end{figure}

\begin{figure}
    \centering
    \includegraphics[width=0.5\textwidth]{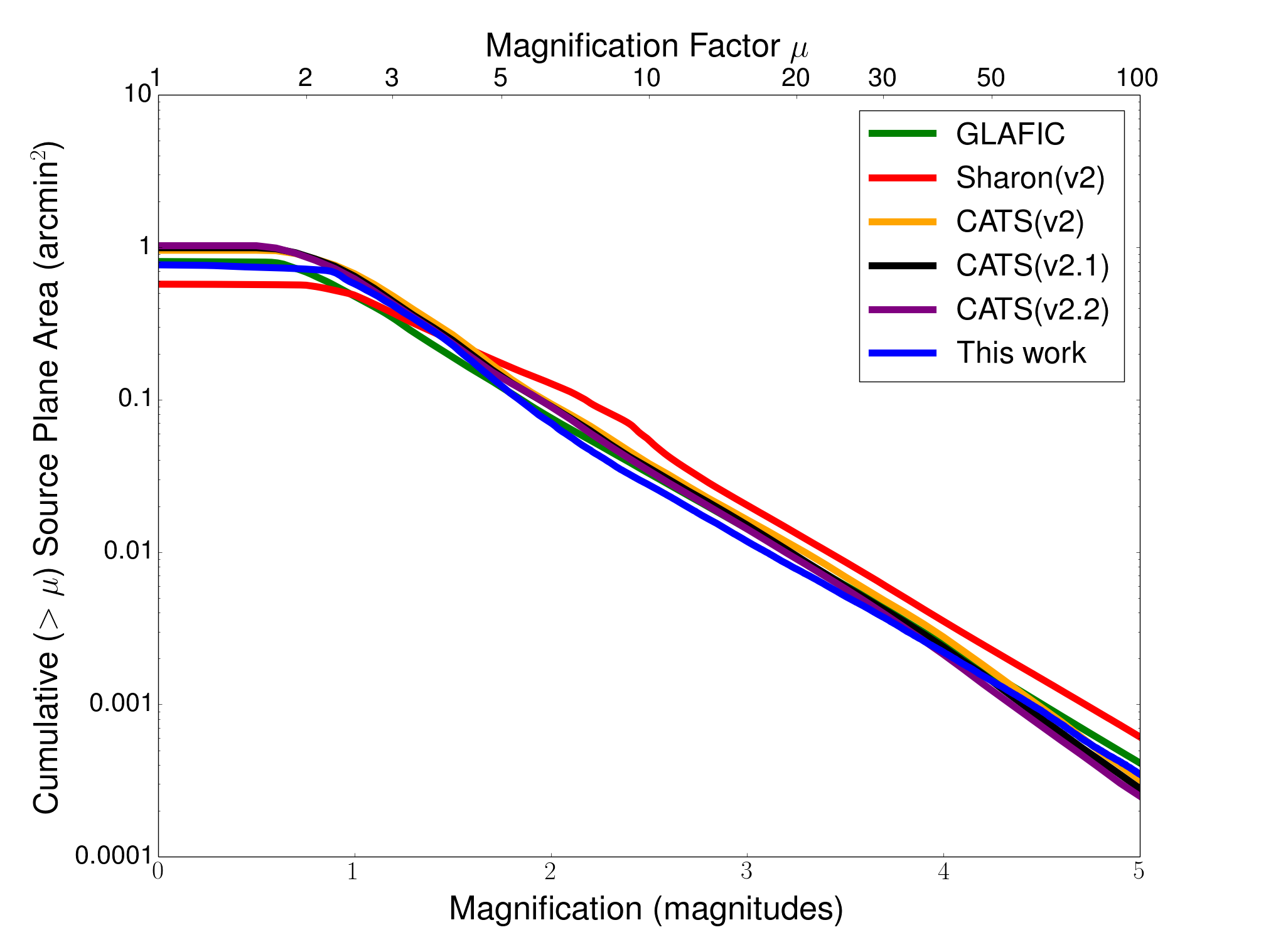}
    \caption{Cumulative source plane area versus magnification at $z=9$. The models used in this comparison are the same as those 
    described in Figure~\ref{fig:surfcompare}.  \label{fig:area}}
\end{figure}

Three teams (\citealp{2014arXiv1409.8663J}, \citealp{2014ApJ...797...98L} and
\citealp{Ish++15}) have published models of \AB{} using new multiple
image constraints identified in the HFF imaging data. Of these teams,
currently only the lens models produced by \citealp{Ish++15} (GLAFIC) are
publicly available through the Mikulski Archive for Space Telescopes
(MAST\footnote{\url{http://archive.stsci.edu/prepds/frontier/lensmodels/}}). We
compare our models to theirs as well as the Sharon v2 models, which
include updated spectroscopic measurements of multiple images
identified before the HFF data were obtained (\citealp{Joh++14}). {Finally, we also compare our model with 
several updates of the CATS(v1) models (Jauzac et al. 2015, private
communication). The CATS(v2) models are presented by \citet{2014arXiv1409.8663J} and use a much larger number of multiple images than we include in our lens model. CATS(v2.1) employs the same set of multiple images as CATS(v2), but makes use of the spectroscopic redshifts obtained in this work. CATS(v2.2) uses the same set of multiple image constraints used in our model.} We compare the surface mass density
profiles (Figure~\ref{fig:surfcompare}) and cumulative magnified source plane areas
(Figure~\ref{fig:area}) predicted by all models described above. The surface mass density profiles agree quite
well at radii where multiple image constraints are plentiful. However,
the models begin to differ rapidly near the boundaries of the
constrained area. Our model disagrees with the CATS models most severely. {It is interesting to note that the three CATS models agree internally extremely well, despite CATS(v2.2) using the same set of multiple images used in this work, a considerably different set than the one used in CATS(v2) and CATS(v2.1).} The significant difference between our model and the CATS models may be due to differences in the modeling techniques or the fact that our method uses additional constraints (weak lensing). Weak lensing constraints have a stronger impact on the model at radii beyond where multiple images are observed. In contrast, there is excellent agreement among the
models in the inferred magnified source plane area. Thus, even though
there may be small but significant differences in the specific details
of each reconstructions, by and large the total integrated properties
are very similar. 

We also note that our model supersedes the model obtained by members
of our team as part of the initial HFF modeling effort based on
pre-HFF data. The uncertainties in this current version of the model
are smaller, consistent with the fact that we have increased
the number of strong lensing constraints.


\begin{figure}
    \centering
    \includegraphics[width=0.5\textwidth]{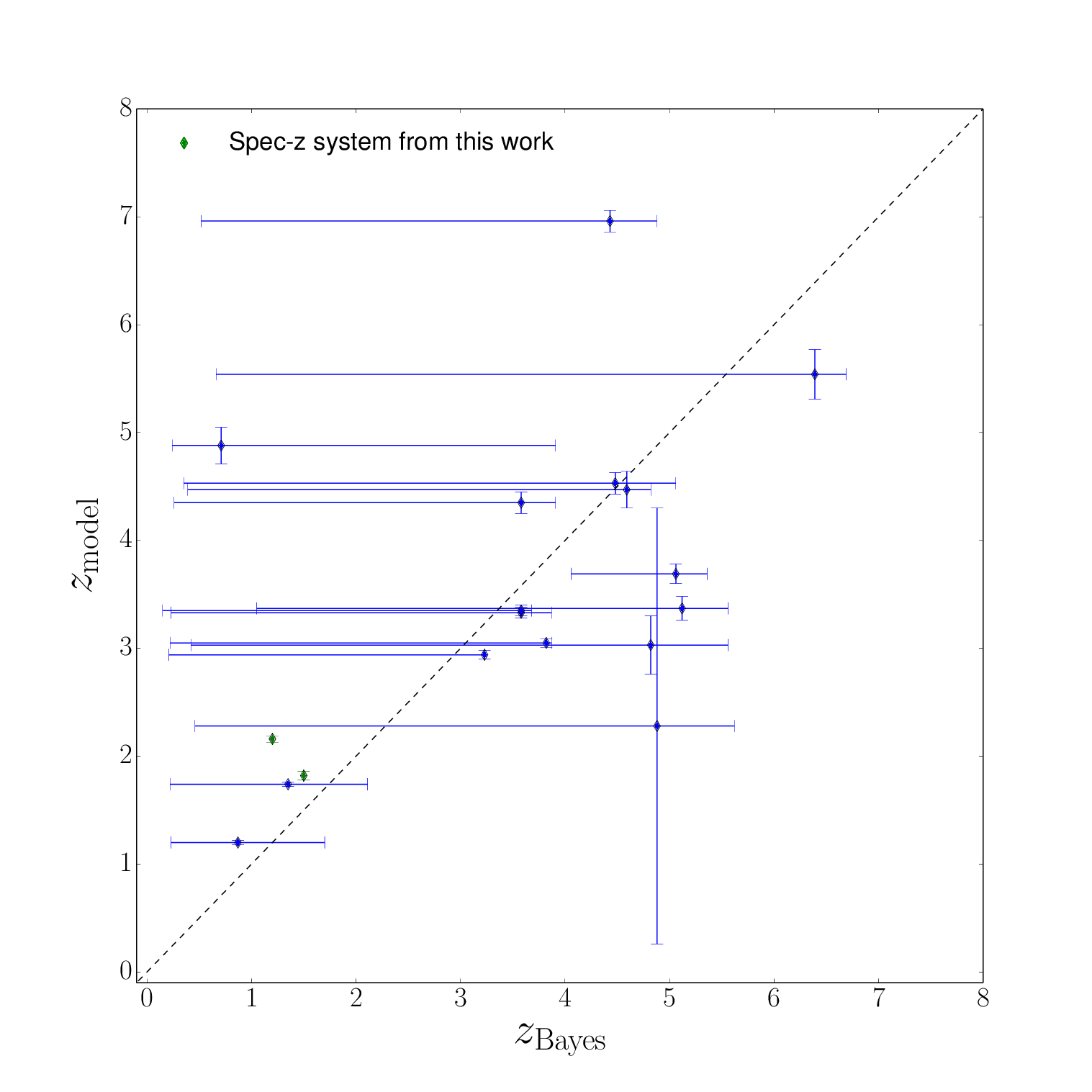}
    \caption{Comparison of the redshifts determined in this work ($z_{\textrm{Bayes}}$) versus the model-predicted redshifts given 
        by \citet{2014arXiv1409.8663J} for all multiple image systems used in the lens model. Note that the previously confirmed 
        spectroscopic systems are left out of this comparison because $z_{\textrm model}$ were not calculated. Two systems (green) 
        are spectroscopically confirmed in this work for the first time and are included in the comparison. For these two objects, 
        we use the new spectroscopic redshift on the horizontal axis in place of $z_{\textrm{Bayes}}$. The $z_{\textrm{model}}$ 
        values are in significant disagreement with the spectroscopic values for these two systems. $z_{\textrm{Bayes}}$ 
        represents the peak of a statistical combination of all available photometric redshift probability density functions 
        \citep{Dahlen:2013p33380}. The vertical error bars reflect 1-$\sigma$ Gaussian error on $z_{\textrm{model}}$, and the 
        horizontal error bars show the $68\%$ credible interval for $z_{\textrm{Bayes}}$. 12/16 systems are consistent at $68\%$ 
    confidence level. }
\label{fig:compare_z}
\end{figure}

We also compare our method of estimating redshifts of multiple image
systems with the one used by the CATS team (\citealp{2014MNRAS.444..268R}, \citealp{2014arXiv1409.8663J}). In 
Figure~\ref{fig:compare_z}, $z_{\textrm{Bayes}}$ is the redshift
obtained from hierarchical Bayesian modeling of the photometric
redshifts obtained in this work. $z_{\textrm{model}}$ is the redshift
obtained by \citet{2014arXiv1409.8663J} by minimizing their analytical
model uncertainty while leaving the redshift as a free parameter. It
is important to check this procedure independently since leaving
$z_{\textrm{model}}$ as a free parameter or predicting additional
multiple images that belong to the same system could in principle lead
to confirmation bias. Overall, we find good agreement between
$z_{\textrm{Bayes}}$ and $z_{\textrm{model}}$, within the admittedly
large uncertainties on $z_{\textrm{Bayes}}$. There are only two new
systems with spectroscopic redshifts available to compare with
$z_{\textrm{model}}$, and they are both inconsistent at $>5\sigma$. This may be due to small
number statistics or perhaps an indication that the uncertainties on
$z_{\textrm{model}}$ are underestimated. More spectroscopic redshifts
are needed to perform this test in a more stringent manner.

\section{The spatial distribution of stellar and dark matter}

\subsection{Stellar mass map}

The \Spitzer{} IRAC $3.6\,\mu$m image samples close to rest-frame
$K$-band for the cluster, so we use the $3.6\,\mu$m fluxes from
cluster members to approximate the cluster stellar mass
distribution. We first selected the red sequence cluster members
brighter than the 25th mag in F814W from the color-magnitude and
color-color diagrams following the procedure described in
\cite{2014MNRAS.444..268R}. We also cross-matched the selected cluster
members with the spectroscopic redshift catalog given by
\citet{Owers:2011ez} to ensure that we included all the cluster
members confirmed with spectroscopy.  We selected a total of 190
bright cluster members for their stellar mass distribution.

To create an image with $3.6\,\mu$m flux from cluster members only, we
first created a mask with value 1 for pixels that belong to cluster
members in the F160W image and 0 otherwise. We then convolved the mask
with the $3.6\,\mu$m PSF to match the IRAC angular resolution, set the
pixels below 10\% of the peak value to zero, and resample the mask
onto the IRAC pixel grid. We obtained the $3.6\,\mu$m map of cluster
members by setting all IRAC pixels not belonging to cluster members to
zero and smoothed the final surface brightness map with a two-pixel
wide Gaussian kernel.

The IRAC surface brightness map was transformed into a surface mass
density map by transforming the 3.6$\mu$m flux into K-band luminosity
and then by multiplying by stellar mass to light ratio derived by
\cite{B+d01} using the so-called ``diet''-Salpeter stellar initial 
mass function (IMF). The resulting stellar mass map in show in the
left panel of Figure~\ref{fig:stellarMassMap}.

The main source of uncertainty on the stellar mass density is the
unknown initial mass function. For example, if one were to adopt a
\cite{Sal55} IMF -- as suggested by studies of massive early-type
galaxies, the stellar mass density would increase by a factor of 1.55.

\subsection{Stellar to total mass ratio}

\begin{figure*}[ht]
\plottwo{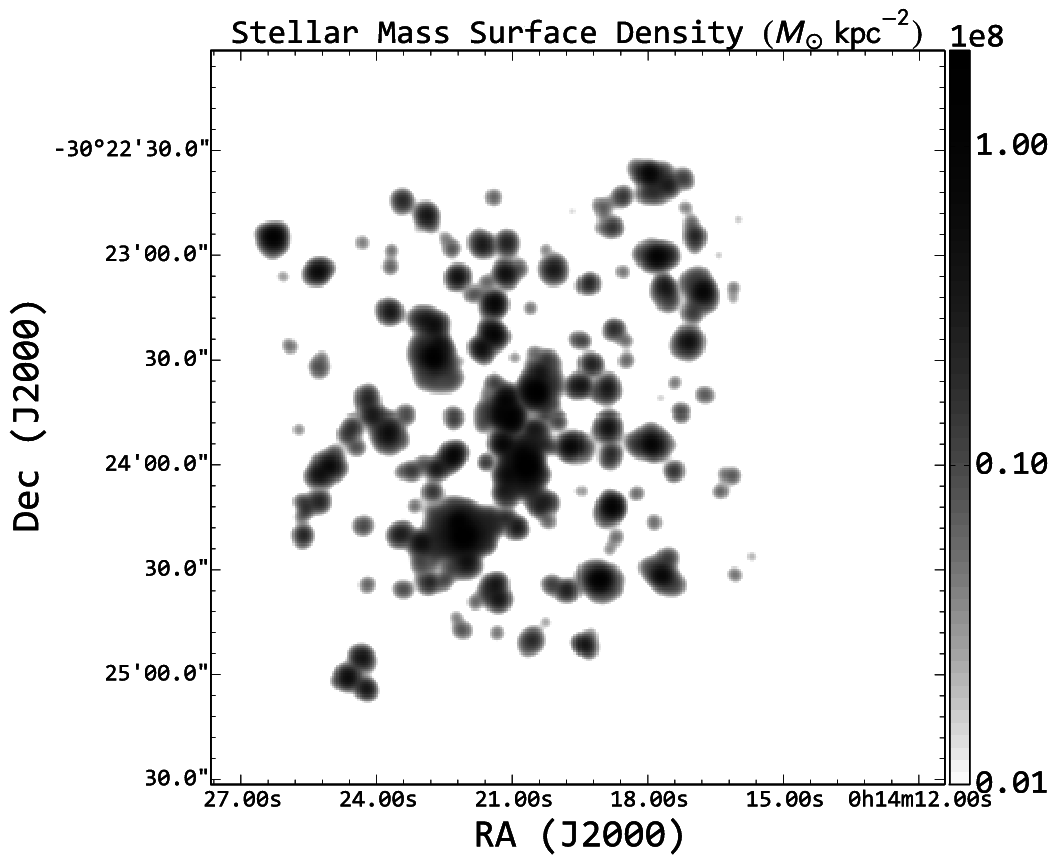}{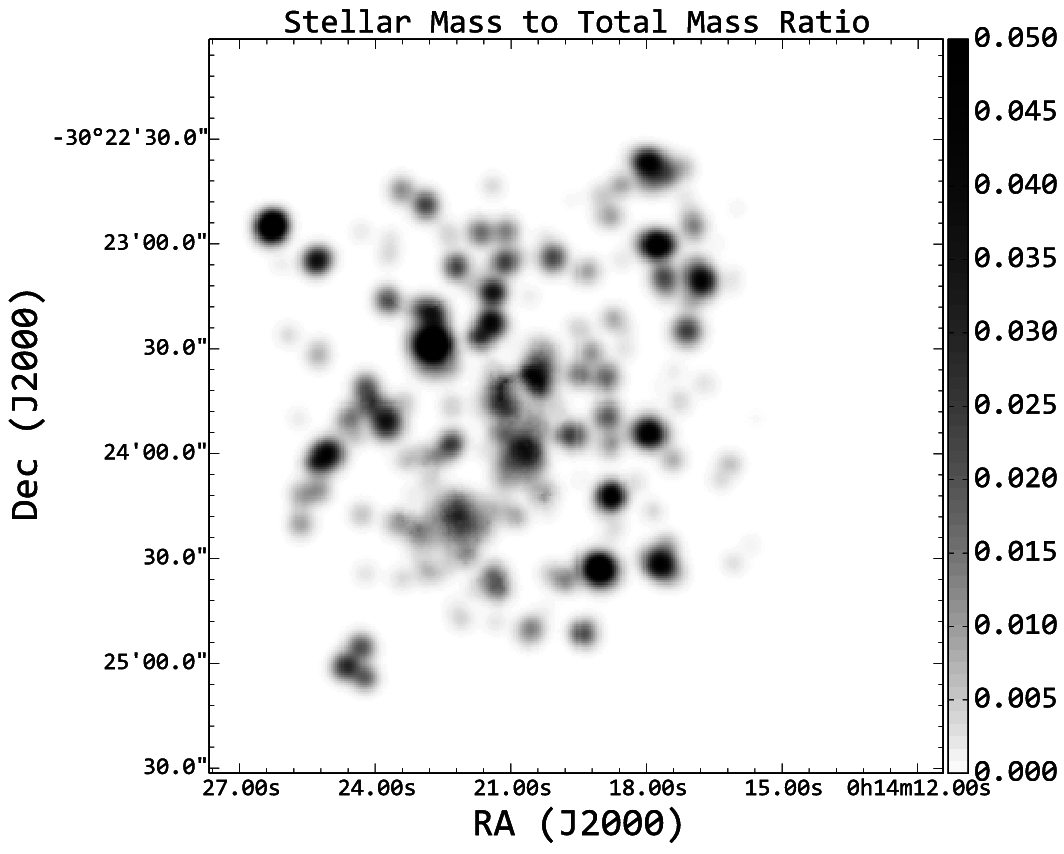}
\caption{Stellar mass surface density (left) and stellar to total mass ratio (right) distributions of \AB{}. Stellar mass surface 
density is in the unit of $M_\odot\,\text{kpc}^{-2}$.  \label{fig:stellarMassMap}}
\end{figure*}

We obtain the stellar to total mass ratio map by dividing the stellar
surface mass density map obtained from photometry by the total surface
mass density map obtained from gravitational lensing. This is shown in
the right panel of Figure~\ref{fig:stellarMassMap}. We note that
resolution effects are non-trivial to take into account, since the
resolution of the lensing map depends on the density of local sources
and the amount of regularization. Thus, the map should be interpreted
keeping in mind this caveat. Interestingly the stellar to total mass
ratio varies significantly across the cluster. Many but not all the
massive ellipticals seem to reach values of 0.05 or more, which are
typical of the central regions of isolated massive galaxies
\cite[e.g.][]{Gav++07}. However, the ratio appears to be significantly lower in the center of the cluster and in the south-east 
quadrant. In future work, we plan to compare the observed map with
those obtained from numerical simulations by taking into account the
effects of finite resolution in the observed mass and light maps, in
order to test whether the spread in stellar to total mass ratio is
reproduced. Furthermore, we plan to carry out a systematic comparison
with mass reconstructions where mass is assumed to follow light up to
a scale factor \citep{Zit++09}. At face value our result is
inconsistent with this assumption for a merging cluster like \AB. However,
uncertainties on both models and resolution effects must be taken into
account in order to evaluate the significance of this apparent
violation. Thus, this result should be considered as preliminary until
confirmed by a more detailed analysis.

\vspace{1cm}

\section{Conclusions}
\label{sec:conc}

In this paper we have used spectroscopic data from the GLASS survey in
combination with ultra-deep imaging data from the HFF program to
construct a strong gravitational lens model for the cluster \AB. In an
effort to obtain a precise and accurate mass model we carried out a
systematic search for spectroscopic redshifts of multiple images and
we applied a rigorous algorithm to select only secure multiple image
systems amongst the dozens that have been proposed in the
literature. The lensing mass map is then combined with a stellar
mass map derived from IRAC photometry to study the relative spatial
distribution of luminous and dark matter.  Our main results can be
summarized as follows:

\begin{enumerate}

\item  We have measured spectroscopic redshifts for \NimgELhiQ{} multiple image systems (quality flag 4 and 3, i.e. secure and 
  probable). We have also confirmed spectroscopically that images 6.1, 6.2, 6.3 belong to the same source. The spectroscopically confirmed images are used to constrain the gravitational lens model. We also obtain 2 tentative redshifts, which are not used to 
  to constrain the mass model, but could potentially be confirmed by future work.

\item From the GLASS data we derive an extensive
redshift catalog of faint emission line systems which we use to test
photometric and lensing determinations of redshift. Generally
speaking, the measurements agree within the 1-$\sigma$ uncertainties,
when nebular emission lines are included in the fitting template.  In
addition, we compare photometric redshifts with redshifts determined
by \citet{2014arXiv1409.8663J}, based on their gravitational lens model and
find an agreement within the large uncertainties of the former. For
the two systems with new spectroscopic redshifts we find a significant
difference with respect to model redshifts. This may be due to small
number statistics or to the model redshifts uncertainties being
underestimated. More spectroscopic redshifts are needed to make a more
stringent test.

\item Our rigorous selection algorithm identifies a total of 25/72 multiple arc systems/images as secure out of a sample of 57/179 
  candidate multiple arc systems/images, compiled from the literature
  and from our own work. Most systems are rejected either on the basis
  of inconsistent morphology or inconsistent spectral energy
  distribution between the candidate multiple images, or because of
  insufficient evidence that they belong to the same source.

\item The derived mass model is found to be very precise, as measured by bootstrap- and redshift-resampling the set of multiple images used as 
  input. Furthermore, we tested how well our model reproduces the magnification of the background SN Ia Tomas (\citealp{Rod++15}). The SN Ia was not used as a constraint to the model and yet its magnification is consistent 
  ($2.03 \pm0.29$ for the supernova vs 2.23$^{+0.08}_{-0.07}$ from our mass model). 

\item \AB{} is confirmed to be an excellent gravitational telescope, with a source plane area of $\sim0.7$ arcminute square being 
  magnified by a factor of 2.

\item We construct a stellar surface mass density map and the stellar to total mass ratio by selecting the light associated with red sequence cluster galaxies and using the total mass density map obtained from strong lensing. Albeit with significant uncertainties, we find that the stellar to mass ratio varies significantly across the cluster, tentatively suggesting that stellar mass does not trace total mass in this interacting system.

\end{enumerate}

\medskip

\acknowledgments

This paper is based on observations made with the NASA/ESA Hubble
Space Telescope, obtained at STScI. We acknowledge support through
grants HST-GO-13459, HST-GO-13177, HST-AR-13235.  We thank the
anonymous referee for the useful comments that help improve the
presentation of this paper.  This work utilizes gravitational lensing
models produced by P.I.s Brada\v{c}, Ebeling, Merten, Zitrin, Sharon,
and Williams funded as part of the \HST\ Frontier Fields program
conducted by STScI. MB, KH, and AH acknowledge support for this work
through a
\Spitzer{} award issued by JPL/Caltech. AH acknowledges support by
NASA Headquarters under the NASA Earth and Space Science Fellowship
Program - Grant ASTRO14F- 0007. MB and AH also acknowledge support
from the special funding as part of the \HST{} Frontier Fields program
conducted by STScI. STScI is operated by AURA, Inc. under NASA
contract NAS 5-26555. The data were obtained from the Mikulski Archive
for Space Telescopes (MAST).  TT acknowledges support by the Packard
Foundation through a Packard Research Fellowship, and thanks the
Osservatorio Astronomico di Monteporzio Catone and the American
Academy in Rome for their kind hospitality during the writing of this
manuscript. BV acknowledges the support from the World Premier
International Research Center Initiative (WPI), MEXT, Japan and the
Kakenhi Grant-in-Aid for Young Scientists (B)(26870140) from the Japan
Society for the Promotion of Science (JSPS). XW acknowledges Michael Maseda for the helpful discussions.

\bibliographystyle{apj}
\bibliography{bibtexlibrary}

\begin{thebibliography}{}
\expandafter\ifx\csname natexlab\endcsname\relax\def\natexlab#1{#1}\fi

\bibitem[{{Atek} {et~al.}(2014){Atek}, {Richard}, {Kneib}, {Clement}, {Egami},
  {Ebeling}, {Jauzac}, {Jullo}, {Laporte}, {Limousin}, \&
  {Natarajan}}]{2014ApJ...786...60A}
{Atek}, H., {Richard}, J., {Kneib}, J.-P., {et~al.} 2014, \apj, 786, 60

\bibitem[{{Atek} {et~al.}(2015){Atek}, {Richard}, {Kneib}, {Jauzac},
  {Schaerer}, {Clement}, {Limousin}, {Jullo}, {Natarajan}, {Egami}, \&
  {Ebeling}}]{2015ApJ...800...18A}
---. 2015, \apj, 800, 18

\bibitem[{{Bayliss} {et~al.}(2014){Bayliss}, {Rigby}, {Sharon}, {Wuyts},
  {Florian}, {Gladders}, {Johnson}, \& {Oguri}}]{Bay++14}
{Bayliss}, M.~B., {Rigby}, J.~R., {Sharon}, K., {et~al.} 2014, \apj, 790, 144

\bibitem[{{Bell} \& {de Jong}(2001)}]{B+d01}
{Bell}, E.~F., \& {de Jong}, R.~S. 2001, \apj, 550, 212

\bibitem[{Bertin \& Arnouts(1996)}]{Bertin:1996p12964}
Bertin, E., \& Arnouts, S. 1996, Astronomy and Astrophysics Supplement, 117,
  393

\bibitem[{Brada{\v c} {et~al.}(2012)Brada{\v c}, Vanzella, Hall, Treu, Fontana,
  Gonzalez, Clowe, Zaritsky, Stiavelli, \& Cl{\'e}ment}]{Bradac:2012p28826}
Brada{\v c}, M., Vanzella, E., Hall, N., {et~al.} 2012, The Astrophysical
  Journal Letters, 755, L7

\bibitem[{{Brada{\v c}} {et~al.}(2005){Brada{\v c}}, {Erben}, {Schneider},
  {Hildebrandt}, {Lombardi}, {Schirmer}, {Miralles}, {Clowe}, \&
  {Schindler}}]{bradac05}
{Brada{\v c}}, M., {Erben}, T., {Schneider}, P., {et~al.} 2005, \aap, 437, 49

\bibitem[{{Brada{\v c}} {et~al.}(2006){Brada{\v c}}, {Clowe}, {Gonzalez},
  {Marshall}, {Forman}, {Jones}, {Markevitch}, {Randall}, {Schrabback}, \&
  {Zaritsky}}]{Bra++06}
{Brada{\v c}}, M., {Clowe}, D., {Gonzalez}, A.~H., {et~al.} 2006, \apj, 652,
  937

\bibitem[{{Brada{\v c}} {et~al.}(2009){Brada{\v c}}, {Treu}, {Applegate},
  {Gonzalez}, {Clowe}, {Forman}, {Jones}, {Marshall}, {Schneider}, \&
  {Zaritsky}}]{bradac09}
{Brada{\v c}}, M., {Treu}, T., {Applegate}, D., {et~al.} 2009, \apj, 706, 1201

\bibitem[{{Brada{\v c}} {et~al.}(2014){Brada{\v c}}, {Ryan}, {Casertano},
  {Huang}, {Lemaux}, {Schrabback}, {Gonzalez}, {Allen}, {Cain}, {Gladders},
  {Hall}, {Hildebrandt}, {Hinz}, {von der Linden}, {Lubin}, {Treu}, \&
  {Zaritsky}}]{2014ApJ...785..108B}
{Brada{\v c}}, M., {Ryan}, R., {Casertano}, S., {et~al.} 2014, \apj, 785, 108

\bibitem[{Brammer {et~al.}(2014)Brammer, Pirzkal, McCullough, \&
  MacKenty}]{Brammer:2014p34990}
Brammer, G.~B., Pirzkal, N., McCullough, P.~R., \& MacKenty, J.~W. 2014, STScI
  IRS

\bibitem[{Brammer {et~al.}(2008)Brammer, van Dokkum, \&
  Coppi}]{Brammer:2008p13280}
Brammer, G.~B., van Dokkum, P.~G., \& Coppi, P. 2008, The Astrophysical
  Journal, 686, 1503

\bibitem[{Brammer {et~al.}(2013)Brammer, van Dokkum, Illingworth, Bouwens,
  Labb{\'e}, Franx, Momcheva, \& Oesch}]{Brammer:2013p27911}
Brammer, G.~B., van Dokkum, P.~G., Illingworth, G.~D., {et~al.} 2013, The
  Astrophysical Journal Letters, 765, L2

\bibitem[{Brammer {et~al.}(2012)Brammer, van Dokkum, Franx, Fumagalli, Patel,
  Rix, Skelton, Kriek, Nelson, Schmidt, Bezanson, Cunha, Erb, Fan, Schreiber,
  Illingworth, Labb{\'e}, Leja, Lundgren, Magee, Marchesini, McCarthy,
  Momcheva, Muzzin, Quadri, Steidel, Tal, Wake, Whitaker, \&
  Williams}]{Brammer:2012p12977}
Brammer, G.~B., van Dokkum, P.~G., Franx, M., {et~al.} 2012, The Astrophysical
  Journal Supplement, 200, 13

\bibitem[{{Clowe} {et~al.}(2006){Clowe}, {Brada{\v{c}}}, {Gonzalez},
  {Markevitch}, {Randall}, {Jones}, \& {Zaritsky}}]{clowe06}
{Clowe}, D., {Brada{\v{c}}}, M., {Gonzalez}, A.~H., {et~al.} 2006, \apjl, 648,
  L109

\bibitem[{{Coe} {et~al.}(2015){Coe}, {Bradley}, \&
  {Zitrin}}]{2015ApJ...800...84C}
{Coe}, D., {Bradley}, L., \& {Zitrin}, A. 2015, \apj, 800, 84

\bibitem[{Dahlen {et~al.}(2013)Dahlen, Mobasher, Faber, Ferguson, Barro,
  Finkelstein, Finlator, Fontana, Gruetzbauch, Johnson, Pforr, Salvato,
  Wiklind, Wuyts, Acquaviva, Dickinson, Guo, Huang, Huang, Newman, Bell,
  Conselice, Galametz, Gawiser, Giavalisco, Grogin, Hathi, Kocevski, Koekemoer,
  Koo, Lee, McGrath, Papovich, Peth, Ryan, Somerville, Weiner, \&
  Wilson}]{Dahlen:2013p33380}
Dahlen, T., Mobasher, B., Faber, S.~M., {et~al.} 2013, The Astrophysical
  Journal, 775, 93

\bibitem[{Ferguson {et~al.}(2000)Ferguson, Dickinson, \&
  Williams}]{Ferguson:2000p22537}
Ferguson, H.~C., Dickinson, M., \& Williams, R. 2000, Annu. Rev. Astro.
  Astrophys., 38, 667

\bibitem[{{Gavazzi} {et~al.}(2007){Gavazzi}, {Treu}, {Rhodes}, {Koopmans},
  {Bolton}, {Burles}, {Massey}, \& {Moustakas}}]{Gav++07}
{Gavazzi}, R., {Treu}, T., {Rhodes}, J.~D., {et~al.} 2007, \apj, 667, 176

\bibitem[{{Gonzaga} \& {et al.}(2012)}]{Gonzaga:2014p26307}
{Gonzaga}, S., \& {et al.} 2012, {The DrizzlePac Handbook} (STScI)

\bibitem[{{Guo} {et~al.}(2013){Guo}, {Ferguson}, {Giavalisco}, {Barro},
  {Willner}, {Ashby}, {Dahlen}, {Donley}, {Faber}, {Fontana}, {Galametz},
  {Grazian}, {Huang}, {Kocevski}, {Koekemoer}, {Koo}, {McGrath}, {Peth},
  {Salvato}, {Wuyts}, {Castellano}, {Cooray}, {Dickinson}, {Dunlop}, {Fazio},
  {Gardner}, {Gawiser}, {Grogin}, {Hathi}, {Hsu}, {Lee}, {Lucas}, {Mobasher},
  {Nandra}, {Newman}, \& {van der Wel}}]{Guo+13}
{Guo}, Y., {Ferguson}, H.~C., {Giavalisco}, M., {et~al.} 2013, \apjs, 207, 24

\bibitem[{{Huang} {et~al.}(2015){Huang}, {Brada{\v c}}, {Lemaux}, {Ryan},
  {Hoag}, {Castellano}, {Amor{\'\i}n}, {Fontana}, {Brammer}, {Cain}, {Lubin},
  {Merlin}, {Schmidt}, {Schrabback}, {Treu}, {Gonzalez}, {Von Der Linden}, \&
  {Knight}}]{2015arXiv150402099H}
{Huang}, K.-H., {Brada{\v c}}, M., {Lemaux}, B.~C., {et~al.} 2015, ArXiv
  e-prints, arXiv:1504.02099

\bibitem[{{Ilbert} {et~al.}(2006){Ilbert}, {Arnouts}, {McCracken},
  {Bolzonella}, {Bertin}, {Le F{\`e}vre}, {Mellier}, {Zamorani}, {Pell{\`o}},
  {Iovino}, {Tresse}, {Le Brun}, {Bottini}, {Garilli}, {Maccagni}, {Picat},
  {Scaramella}, {Scodeggio}, {Vettolani}, {Zanichelli}, {Adami}, {Bardelli},
  {Cappi}, {Charlot}, {Ciliegi}, {Contini}, {Cucciati}, {Foucaud}, {Franzetti},
  {Gavignaud}, {Guzzo}, {Marano}, {Marinoni}, {Mazure}, {Meneux}, {Merighi},
  {Paltani}, {Pollo}, {Pozzetti}, {Radovich}, {Zucca}, {Bondi}, {Bongiorno},
  {Busarello}, {de La Torre}, {Gregorini}, {Lamareille}, {Mathez}, {Merluzzi},
  {Ripepi}, {Rizzo}, \& {Vergani}}]{Ilb++06}
{Ilbert}, O., {Arnouts}, S., {McCracken}, H.~J., {et~al.} 2006, \aap, 457, 841

\bibitem[{{Ishigaki} {et~al.}(2015){Ishigaki}, {Kawamata}, {Ouchi}, {Oguri},
  {Shimasaku}, \& {Ono}}]{Ish++15}
{Ishigaki}, M., {Kawamata}, R., {Ouchi}, M., {et~al.} 2015, \apj, 799, 12

\bibitem[{{Jauzac} {et~al.}(2014){Jauzac}, {Richard}, {Jullo}, {Cl{\'e}ment},
  {Limousin}, {Kneib}, {Ebeling}, {Rodney}, {Natarajan}, {Atek}, {Massey},
  {Eckert}, {Egami}, \& {Rexroth}}]{2014arXiv1409.8663J}
{Jauzac}, M., {Richard}, J., {Jullo}, E., {et~al.} 2014, ArXiv e-prints,
  arXiv:1409.8663

\bibitem[{{Johnson} {et~al.}(2014){Johnson}, {Sharon}, {Bayliss}, {Gladders},
  {Coe}, \& {Ebeling}}]{Joh++14}
{Johnson}, T.~L., {Sharon}, K., {Bayliss}, M.~B., {et~al.} 2014, \apj, 797, 48

\bibitem[{Johnson {et~al.}(2014)Johnson, Sharon, Bayliss, Gladders, Coe, \&
  Ebeling}]{Johnson:2014p37801}
Johnson, T.~L., Sharon, K., Bayliss, M.~B., {et~al.} 2014, The Astrophysical
  Journal, 797, 48

\bibitem[{{Kempner} \& {David}(2004)}]{2004MNRAS.349..385K}
{Kempner}, J.~C., \& {David}, L.~P. 2004, \mnras, 349, 385

\bibitem[{{Kneib} \& {Natarajan}(2011)}]{K+N11}
{Kneib}, J.-P., \& {Natarajan}, P. 2011, \aapr, 19, 47

\bibitem[{Koekemoer {et~al.}(2003)Koekemoer, Fruchter, Hook, \&
  Hack}]{Koekemoer:2003p31861}
Koekemoer, A.~M., Fruchter, A.~S., Hook, R.~N., \& Hack, W. 2003, The 2002 HST
  Calibration Workshop : Hubble after the Installation of the ACS and the
  NICMOS Cooling System, 337

\bibitem[{K{\"u}mmel {et~al.}(2011)K{\"u}mmel, Kuntschner, Walsh, \&
  Bushouse}]{Kummel:2011p33451}
K{\"u}mmel, M., Kuntschner, H., Walsh, J.~R., \& Bushouse, H. 2011, ST-ECF
  Instrument Science Report WFC3-2011-01, 1

\bibitem[{{Lam} {et~al.}(2014){Lam}, {Broadhurst}, {Diego}, {Lim}, {Coe},
  {Ford}, \& {Zheng}}]{2014ApJ...797...98L}
{Lam}, D., {Broadhurst}, T., {Diego}, J.~M., {et~al.} 2014, \apj, 797, 98

\bibitem[{{Laporte} {et~al.}(2015){Laporte}, {Streblyanska}, {Kim},
  {Pell{\'o}}, {Bauer}, {Bina}, {Brammer}, {De Leo}, {Infante}, \&
  {P{\'e}rez-Fournon}}]{Lap++15}
{Laporte}, N., {Streblyanska}, A., {Kim}, S., {et~al.} 2015, \aap, 575, A92

\bibitem[{{Merten} {et~al.}(2011){Merten}, {Coe}, {Dupke}, {Massey}, {Zitrin},
  {Cypriano}, {Okabe}, {Frye}, {Braglia}, {Jim{\'e}nez-Teja}, {Ben{\'{\i}}tez},
  {Broadhurst}, {Rhodes}, {Meneghetti}, {Moustakas}, {Sodr{\'e}}, {Krick}, \&
  {Bregman}}]{2011MNRAS.417..333M}
{Merten}, J., {Coe}, D., {Dupke}, R., {et~al.} 2011, \mnras, 417, 333

\bibitem[{{Merten} {et~al.}(2015){Merten}, {Meneghetti}, {Postman}, {Umetsu},
  {Zitrin}, {Medezinski}, {Nonino}, {Koekemoer}, {Melchior}, {Gruen},
  {Moustakas}, {Bartelmann}, {Host}, {Donahue}, {Coe}, {Molino}, {Jouvel},
  {Monna}, {Seitz}, {Czakon}, {Lemze}, {Sayers}, {Balestra}, {Rosati},
  {Ben{\'{\i}}tez}, {Biviano}, {Bouwens}, {Bradley}, {Broadhurst}, {Carrasco},
  {Ford}, {Grillo}, {Infante}, {Kelson}, {Lahav}, {Massey}, {Moustakas},
  {Rasia}, {Rhodes}, {Vega}, \& {Zheng}}]{2015ApJ...806....4M}
{Merten}, J., {Meneghetti}, M., {Postman}, M., {et~al.} 2015, \apj, 806, 4

\bibitem[{{Newman} {et~al.}(2013){Newman}, {Treu}, {Ellis}, \&
  {Sand}}]{New++13}
{Newman}, A.~B., {Treu}, T., {Ellis}, R.~S., \& {Sand}, D.~J. 2013, \apj, 765,
  25

\bibitem[{{Oke}(1974)}]{Oke74}
{Oke}, J.~B. 1974, \apjs, 27, 21

\bibitem[{Owers {et~al.}(2011)Owers, Randall, Nulsen, Couch, David, \&
  Kempner}]{Owers:2011ez}
Owers, M.~S., Randall, S.~W., Nulsen, P. E.~J., {et~al.} 2011, The
  Astrophysical Journal, 728, 27

\bibitem[{{Pettini} {et~al.}(2002){Pettini}, {Rix}, {Steidel}, {Adelberger},
  {Hunt}, \& {Shapley}}]{Pet++02}
{Pettini}, M., {Rix}, S.~A., {Steidel}, C.~C., {et~al.} 2002, \apj, 569, 742

\bibitem[{Postman {et~al.}(2012)Postman, Coe, Ben{\'\i}tez, Bradley,
  Broadhurst, Donahue, Ford, Graur, Graves, Jouvel, Koekemoer, Lemze,
  Medezinski, Molino, Moustakas, Ogaz, Riess, Rodney, Rosati, Umetsu, Zheng,
  Zitrin, Bartelmann, Bouwens, Czakon, Golwala, Host, Infante, Jha,
  Jimenez-Teja, Kelson, Lahav, Lazkoz, Maoz, McCully, Melchior, Meneghetti,
  Merten, Moustakas, Nonino, Patel, Reg{\"o}s, Sayers, Seitz, \&
  Wel}]{Postman:2012p27556}
Postman, M., Coe, D., Ben{\'\i}tez, N., {et~al.} 2012, The Astrophysical
  Journal Supplement, 199, 25

\bibitem[{{Richard} {et~al.}(2014){Richard}, {Jauzac}, {Limousin}, {Jullo},
  {Cl{\'e}ment}, {Ebeling}, {Kneib}, {Atek}, {Natarajan}, {Egami}, {Livermore},
  \& {Bower}}]{2014MNRAS.444..268R}
{Richard}, J., {Jauzac}, M., {Limousin}, M., {et~al.} 2014, \mnras, 444, 268

\bibitem[{{Rodney} {et~al.}(2015){Rodney}, {Patel}, {Scolnic}, {Foley},
  {Molino}, {Brammer}, {Jauzac}, {Bradac}, {Coe}, {Broadhurst}, {Diego},
  {Graur}, {Hjorth}, {Hoag}, {Jha}, {Johnson}, {Kelly}, {Lam}, {McCully},
  {Medezinski}, {Meneghetti}, {Merten}, {Richard}, {Riess}, {Sharon},
  {Strolger}, {Treu}, {Wang}, {Williams}, \& {Zitrin}}]{Rod++15}
{Rodney}, S.~A., {Patel}, B., {Scolnic}, D., {et~al.} 2015, ArXiv e-prints,
  arXiv:1505.06211

\bibitem[{{Ryan} {et~al.}(2014){Ryan}, {Gonzalez}, {Lemaux}, {Brada{\v c}},
  {Casertano}, {Allen}, {Cain}, {Gladders}, {Hall}, {Hildebradt}, {Hinz},
  {Huang}, {Lubin}, {Schrabback}, {Stiavelli}, {Treu}, {von der Linden}, \&
  {Zaritsky}}]{2014ApJ...786L...4R}
{Ryan}, Jr., R.~E., {Gonzalez}, A.~H., {Lemaux}, B.~C., {et~al.} 2014, \apjl,
  786, L4

\bibitem[{{Salpeter}(1955)}]{Sal55}
{Salpeter}, E.~E. 1955, \apj, 121, 161

\bibitem[{{Sand} {et~al.}(2008){Sand}, {Treu}, {Ellis}, {Smith}, \&
  {Kneib}}]{San++08}
{Sand}, D.~J., {Treu}, T., {Ellis}, R.~S., {Smith}, G.~P., \& {Kneib}, J. 2008,
  \apj, 674, 711

\bibitem[{{Sharon} {et~al.}(2014){Sharon}, {Gladders}, {Rigby}, {Wuyts},
  {Bayliss}, {Johnson}, {Florian}, \& {Dahle}}]{Sha++14}
{Sharon}, K., {Gladders}, M.~D., {Rigby}, J.~R., {et~al.} 2014, \apj, 795, 50

\bibitem[{{Shirazi} {et~al.}(2014){Shirazi}, {Vegetti}, {Nesvadba}, {Allam},
  {Brinchmann}, \& {Tucker}}]{Shi++14}
{Shirazi}, M., {Vegetti}, S., {Nesvadba}, N., {et~al.} 2014, \mnras, 440, 2201

\bibitem[{Treu {et~al.}(2015)Treu, Schmidt, Brada{\v c}, Brammer, Dijkstra,
  Dressler, Fontana, Gavazzi, Henry, Hoag, Jones, Kelly, Malkan, Mason,
  Pentericci, Poggianti, Stiavelli, Trenti, von~der Linden, Vulcani, \&
  Wang}]{Treu:2015p36793}
Treu, T., Schmidt, K.~B., Brada{\v c}, M., {et~al.} 2015, apJ, submitted.

\bibitem[{{Williams} {et~al.}(1996){Williams}, {Blacker}, {Dickinson}, {Dixon},
  {Ferguson}, {Fruchter}, {Giavalisco}, {Gilliland}, {Heyer}, {Katsanis},
  {Levay}, {Lucas}, {McElroy}, {Petro}, {Postman}, {Adorf}, \&
  {Hook}}]{Wil++96}
{Williams}, R.~E., {Blacker}, B., {Dickinson}, M., {et~al.} 1996, \aj, 112,
  1335

\bibitem[{{Yee} {et~al.}(1996){Yee}, {Ellingson}, {Bechtold}, {Carlberg}, \&
  {Cuillandre}}]{Yee++96}
{Yee}, H.~K.~C., {Ellingson}, E., {Bechtold}, J., {Carlberg}, R.~G., \&
  {Cuillandre}, J.-C. 1996, \aj, 111, 1783

\bibitem[{{Zitrin} {et~al.}(2009){Zitrin}, {Broadhurst}, {Umetsu}, {Coe},
  {Ben{\'{\i}}tez}, {Ascaso}, {Bradley}, {Ford}, {Jee}, {Medezinski},
  {Rephaeli}, \& {Zheng}}]{Zit++09}
{Zitrin}, A., {Broadhurst}, T., {Umetsu}, K., {et~al.} 2009, \mnras, 396, 1985

\bibitem[{{Zitrin} {et~al.}(2014){Zitrin}, {Zheng}, {Broadhurst}, {Moustakas},
  {Lam}, {Shu}, {Huang}, {Diego}, {Ford}, {Lim}, {Bauer}, {Infante}, {Kelson},
  \& {Molino}}]{2014ApJ...793L..12Z}
{Zitrin}, A., {Zheng}, W., {Broadhurst}, T., {et~al.} 2014, \apjl, 793, L12

\end{thebibliography}

\clearpage
\LongTables

\newcolumntype{L}[1]{>{\raggedright\let\newline\\\arraybackslash\hspace{5pt}}m{#1}}
\tabletypesize{\scriptsize} \tabcolsep=0.08cm
\begin{deluxetable*}{L{4em}ccccccccccccc} \tablecolumns{13}
\tablewidth{0pt}
\tablecaption{Multiply lensed arc systems identified in the \AB{} field\label{tab:mult_images}}
\tablehead{
  \colhead{ID$_\textrm{arc}$\tablenotemark{1}} &
  \colhead{R.A.} &
  \colhead{Dec.} &
  \colhead{F140W} &
  \colhead{$z_{\textrm{spec}}$} &
  \colhead{$z_{\textrm{Bayes}}$\tablenotemark{2}} &
  \multicolumn{4}{c}{Quality} &
  \multicolumn{2}{c}{Pass Flag} &
  \colhead{Magnification\tablenotemark{4}} \\
  & (deg.) & (deg.) & (mag) & & & \multicolumn{4}{c}{\hrulefill} & \multicolumn{2}{c}{\hrulefill} & \\
  & & & & & & $z_{\textrm{Bayes}}$\tablenotemark{3} & Color-$\chi^2$ & Morphology & Contaminated? & System & Image & \\
}
\startdata
1.1     J  &  3.597540  &   -30.403920  & 22.16 &        & 1.70 [1.08-2.01] &  & 0.487 & 8 & 0 & 1 & 1 & 3.88 [4.04-5.26] \\ 
1.2     J  &  3.595960  &   -30.406820  & 22.24 &        &         &  & 3.079 & 8 & 0 &    & 1 & 3.81 [3.63-4.64] \\ 
1.3     G  &  3.586210  &   -30.409990  & 22.92 &  1.50  &         &  & 2.572 & 8 & 0 &    & 1 & 3.73 [3.52-4.66] \\ 
\hline\noalign{\smallskip}
2.1     J  &  3.583250  &   -30.403350  & 24.87 &  1.20  & 0.01 &  & 0.892 & 8 & 0 & 1 & 1 & 7.34 [7.49-9.18] \\ 
2.2\tablenotemark{5}     J  &  3.597290  &   -30.396720  & \nodata &      &         &  & \nodata & 8 & 0 &    & 1 & 2.53 [2.45-2.59] \\ 
2.3     J  &  3.585420  &   -30.399900  & 27.46 &        &         &  & 1.702 & 8 & 1 &    & 1 & 14.84 [9.88-17.08] \\ 
2.4     J  &  3.586420  &   -30.402130  & 26.34 &        &         &  & 0.696 & 8 & 1 &    & 1 & 4.08 [4.22-4.66] \\ 
\hline\noalign{\smallskip}
3.1     T  &  3.589380  &   -30.393870  & 24.28 &  3.98  & 4.06 [3.82-4.38] &  & 0.830 & 8 & 0 & 1 & 1 & 16.59 [10.17-23.33] \\
\enddata
\end{deluxetable*}

\clearpage
\LongTables
\begin{landscape}

\tabletypesize{\scriptsize} \tabcolsep=0.11cm
\begin{deluxetable*}{lcccccccccccc} \tablecolumns{13}
\tablewidth{0pt}
\tablecaption{Emission line detection results on multiply and singly lensed sources}
\tablehead{
 \colhead{ID$_\textrm{GLASS}$} &
 \colhead{ID$_\textrm{arc}$} &
 \colhead{R.A.} &
 \colhead{Dec.} &
 \colhead{F140W} &
 \colhead{$z_\textrm{phot}$} &
 \colhead{$z_\textrm{spec}$} &
 \colhead{Quality} &
 \colhead{P.A.} &
 \colhead{N$_\mathrm{lines}$} &
 \colhead{Line(s)} &
 \colhead{Line Flux} &
 \colhead{Magnification\tablenotemark{1}} \\
 \colhead{} &
 \colhead{} &
 \colhead{(deg.)} &
 \colhead{(deg.)} &
 \colhead{(mag)} &
 \colhead{} &
 \colhead{} &
 \colhead{} &
 \colhead{(deg.)} &
 \colhead{} &
 \colhead{} &
 \colhead{(10$^{-17}$\cgs)} &
 \colhead{}
}
\startdata
\multirow{2}{*}{00336}  &    \multirow{2}{*}{1.3}   &  \multirow{2}{*}{3.586232335}  &  \multirow{2}{*}{-30.409968879}   & \multirow{2}{*}{22.92$\pm$0.01}  & \multirow{2}{*}{ $3.54^{+0.05}_{-3.24}$ } &  \multirow{2}{*}{1.50}  &  \multirow{2}{*}{3}  &  135 &  3  &  [OII] [OIII] H$\alpha$  &  4.1$\pm$1.0 10.4$\pm$0.9 1.5$\pm$0.5  &\multirow{2}{*}{3.74 [3.63-4.22]} \\
 &  &  &  &  &  &  &  &  233  &  1  &  H$\alpha$ &  0.3$\pm$0.5  &  \\ 
\multirow{2}{*}{00433}  &    \multirow{2}{*}{6.3}   &  \multirow{2}{*}{3.586426870}  &  \multirow{2}{*}{-30.409349590}   & \multirow{2}{*}{24.29$\pm$0.02}  & \multirow{2}{*}{ $1.79^{+0.48}_{-1.15}$ } &  \multirow{2}{*}{2.02}  &  \multirow{2}{*}{4}  &  135 &  2  &  [OII] [OIII]            &  0.4$\pm$1.0 10.1$\pm$0.7  &\multirow{2}{*}{5.32 [5.08-6.09]} \\
 &  &  &  &  &  &  &  &  233  &  1  &  [OIII]    &  7.0$\pm$0.6 &  \\ 
\multirow{2}{*}{00523 } &    \multirow{2}{*}{6.2}   &  \multirow{2}{*}{3.594060460}  &  \multirow{2}{*}{-30.407997680}   & \multirow{2}{*}{24.66$\pm$0.02}  & \multirow{2}{*}{ $0.05^{+3.73}_{-0.01}$ } &  \multirow{2}{*}{2.02}  &  \multirow{2}{*}{4}  &  135 &  2  &  [OII] [OIII]            &  2.4$\pm$1.1 5.3$\pm$0.6 &\multirow{2}{*}{1.75 [1.86-2.09]} \\
 &  &  &  &  &  &  &  &  233  &  2  &  [OII] [OIII]  &  0.7$\pm$0.9 7.5$\pm$0.6  & \\ 
\multirow{2}{*}{00807}  &   \multirow{2}{*}{22.2}   &  \multirow{2}{*}{3.600055342}  &  \multirow{2}{*}{-30.404393062}   & \multirow{2}{*}{27.18$\pm$0.15}  & \multirow{2}{*}{ $1.65^{+0.12}_{-0.20}$ } &  \multirow{2}{*}{4.84 } &  \multirow{2}{*}{2}  &  135 &  1  &  MgII                    &  1.3$\pm$0.5  &\multirow{2}{*}{5.10 [4.83-6.16]} \\
 &  &  &  &  &  &  &  &  233  &  1  &   MgII  &  0.7$\pm$0.4  & \\
\enddata
\tablecomments{The first part of this table consists of emission line identifications for the arcs of quality levels 4, 3, and 2, whereas the
second part is comprised of only high-confidence (quality 3 or 4) emission line objects newly discovered during the blind search
procedure, as described in Section~\ref{subsec:blind}. The uncertainty ranges and error bars quoted here all represent 1-$\sigma$ confidence level.}
\tablenotetext{1}{Magnifications of multiply-lensed objects are calculated assuming redshift $z_\textrm{spec}$, which was only used in the lens model for quality$>2$ objects. }
\label{tab:ELtot}
\end{deluxetable*}

\clearpage
\end{landscape}

\appendix
\label{appendix1}

\begin{figure*}[ht]
\centering
\plottwo{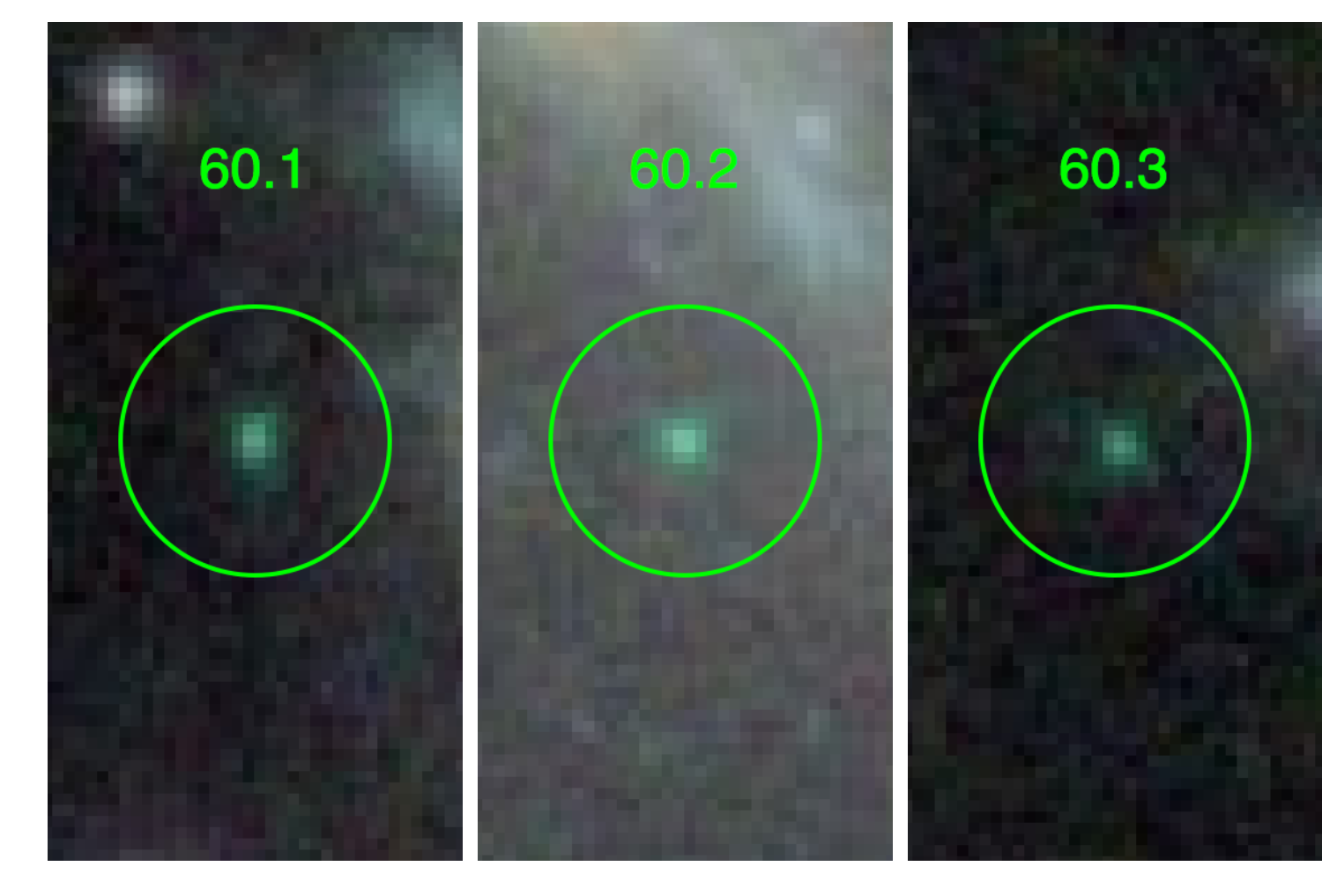}{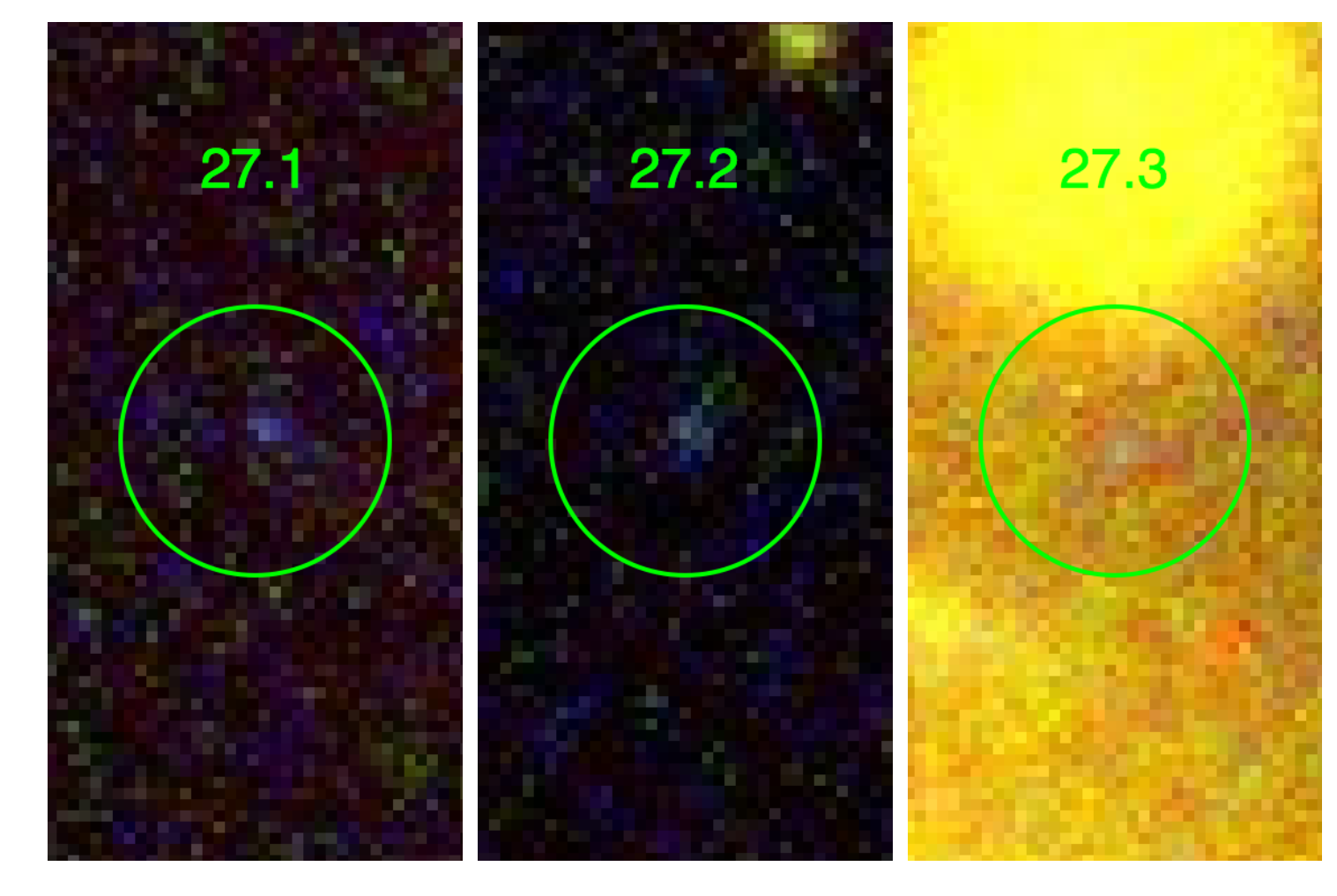}
\caption{\textbf{Left}: System 60, an example of a multiple image system whose images fulfill the morphology criteria, all three 
of which receive maximum grades ($M$) of 8. The images all possess the same point-like morphology and have similar surface 
brightnesses. This multiple image system is discovered for the first time in this work. Unfortunately, the photometric redshifts 
for this system were not reliable, so we could not use the image system in the lens model. The RGB image for system 60 is made 
from a combination of frames: F105W, F125W and F160W, with a pixel scale of 60 milli-arcseconds. \textbf{Right}: System 27, a 
multiple image system whose images fail to fulfill the color/morphology criteria due to poor morphology grades. All three images 
receive a morphology grade of 5. {Both inspectors independently gave the images a low grade because they were too faint to 
discern any shareable features. The color scale is the same for all three images, showing the severe foreground contamination due 
to the ICL in image 27.3. Without spectroscopic redshifts and morphological details, there is no assurance that these images are 
of the same source other than their colors, which are largely uncertain due to the extreme faintness of the images. The RGB image 
for system 27 is made from a combination of frames: F435W, F814W and F125W, with a pixel scale of 60 milli-arcseconds. Out of all 
possible RGB combinations of HST images, the images are most discernible in this one. In both system 60 and system 27, the radius 
of the green circles is 0.72 arcseconds.}}
\label{fig:morph}
\end{figure*}

\begin{figure*}[ht]
\centering
\plottwo{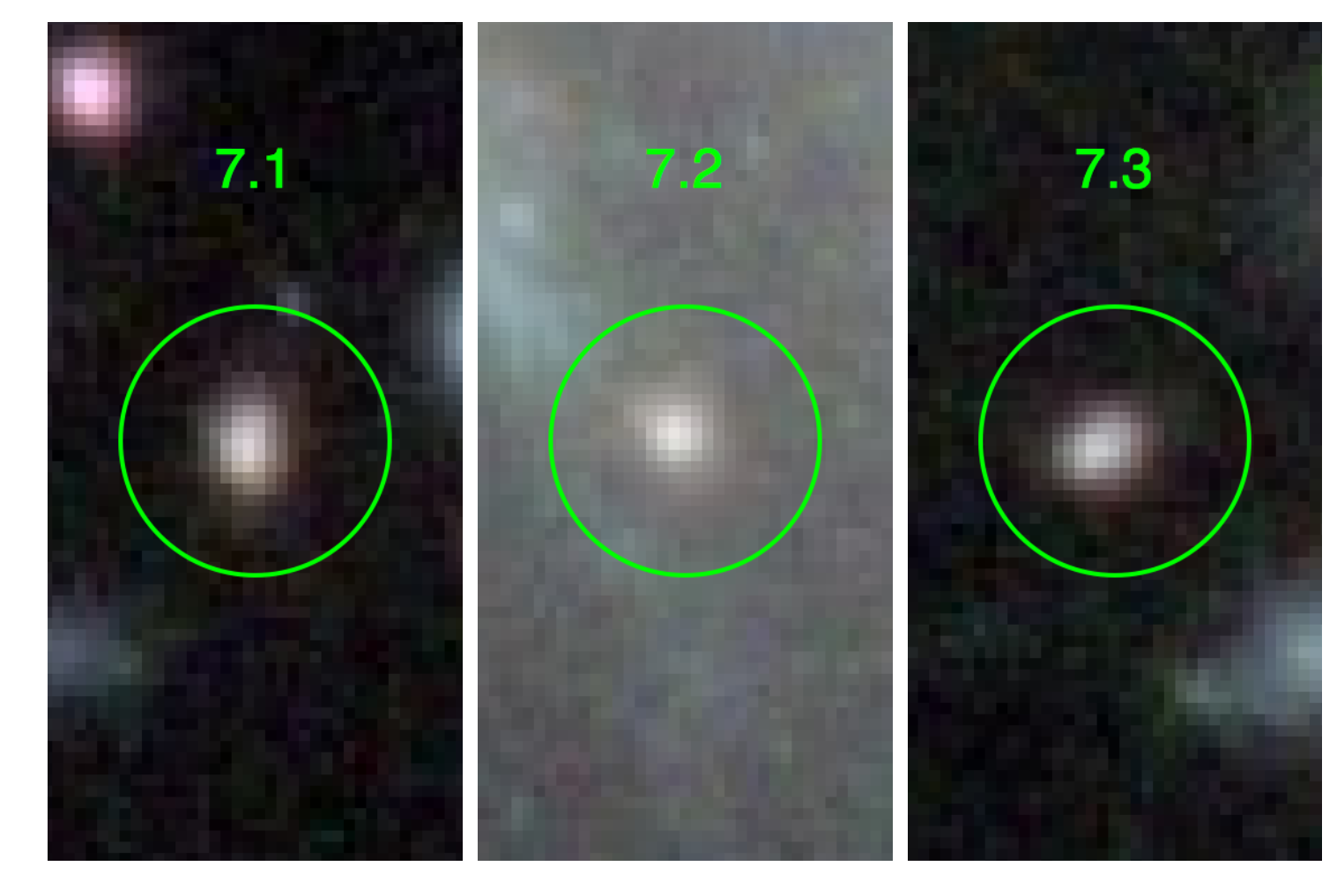}{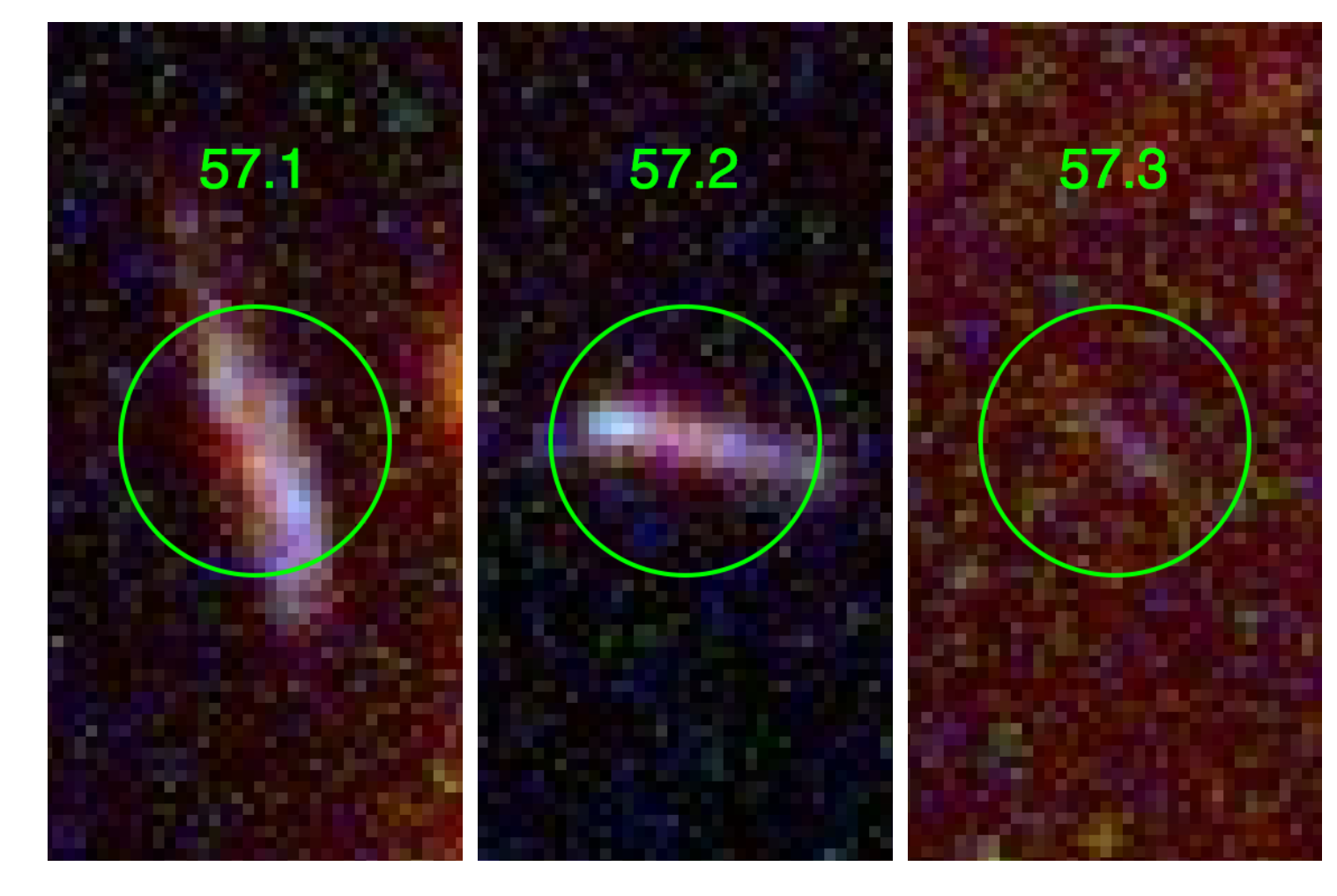}
\caption{\textbf{Left}: System 7, an example of a multiple image system whose images receive high morphology grades ($M$) of 8, 7 
and 8, respectively. The RGB image for system 7  is made from a combination of frames: F105W, F125W and F160W. \textbf{Right}: 
System 57, {a multiple image system that we can rule out due to an impossible lensing configuration, for which $M=0$ is 
assigned for all images. The proposed images are in a naked-cusp configuration and should all therefore have similar 
magnification. 57.3 is barely detected, yet 57.1 and 57.2 are detected at high S/N, thus invalidating the system. If 57.3 were 
simply misidentified, there should still be a counter-image with the same color, similar magnification and elongation as 57.1 and 
57.2 nearby 57.3. We find no such counter-image. There may be another lensing configuration in which 57.1 and 57.2 are indeed 
multiple images of the same source, in which case there may be additional counter-images elsewhere in the cluster field that have 
yet to be found. The RGB image for system 57 is made from a combination of frames: F435W, F606W and F105W. In both systems 7 and 
57, the radius of the green circles is 0.72 arcseconds.} }
\label{fig:morph}
\end{figure*}

\begin{figure*}[ht]
\centering
\plottwo{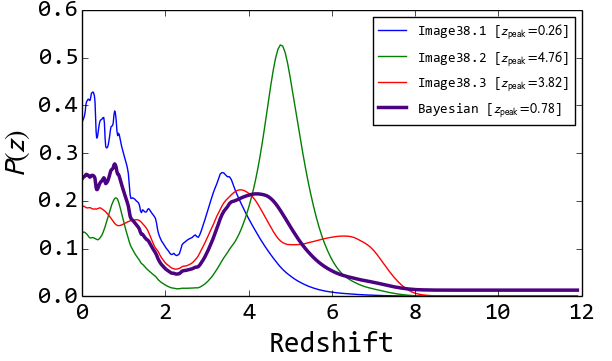}{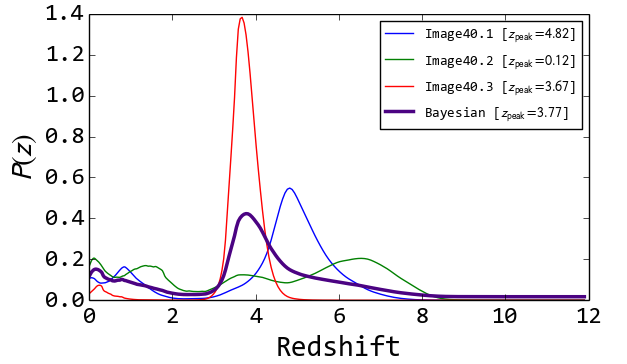}
\caption{Two examples of arc systems that we exclude from the lens model due to poorly-constrained redshifts. Shown are the 
photometric redshift probability density functions, $P(z)$, for the individual images (blue, green, red) along with a combined 
$P(z)$ obtained via hierarchical Bayesian modeling (purple). \textbf{Left} System 38 is excluded because it is bimodal with 
comparable probabilities under each peak. \textbf{Right} System 40 is excluded because the individual photometric redshifts of 
images 40.1, 40.2 and 40.3 are in significant disagreement. }
\end{figure*} 

\begin{figure*}[ht]
\centering
\plottwo{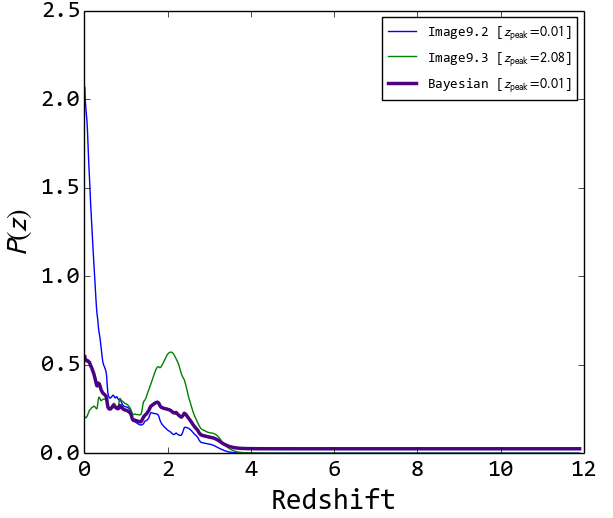}{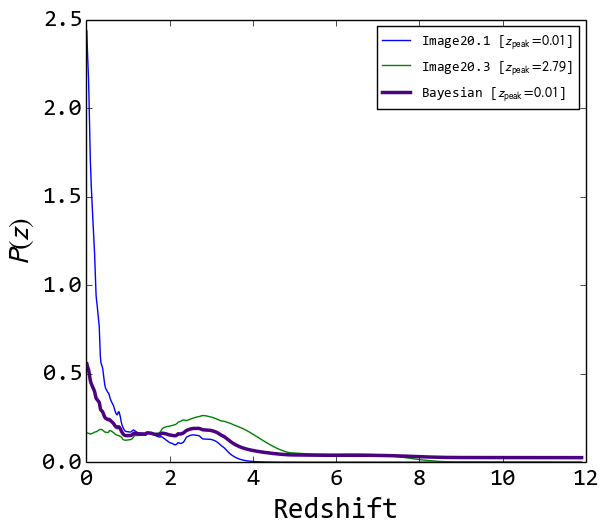}
\caption{Two examples of arc systems that we exclude from the lens model due to a peak in the combined Bayesian redshift below the cluster redshift of \AB, $z=0.308$. \textbf{Left} System 9. The $P(z)$ for image 9.1 is not shown because the object is flagged as contaminated. \textbf{Right} System 20. Likewise, $P(z)$ for image 20.2 is not shown because the object is flagged as contaminated. }
\end{figure*} 


\end{document}